\documentclass[preprint,12pt]{elsarticle}

\usepackage{amssymb}
\usepackage{amsmath}
\usepackage{graphicx}
\usepackage{multirow}
\usepackage[leftcaption]{sidecap}
\usepackage{amsmath}
\usepackage{amssymb}
\usepackage{color}
\usepackage{enumitem}
\usepackage{verbatim}
\usepackage{amssymb,amsfonts}
\usepackage{algorithmic}
\usepackage{textcomp}
\usepackage{subfigure}
\usepackage{subcaption}
\usepackage{setspace}
\usepackage{enumitem}
\usepackage{algorithm}
\usepackage{bbding}
\usepackage{framed}
\usepackage{textcomp}
\usepackage[flushleft]{threeparttable}
\usepackage{booktabs,tabularx,tabu}
\usepackage{gensymb}
\usepackage{lipsum}
\usepackage{dblfloatfix}
\usepackage{amsmath}
\usepackage{xcolor}

\usepackage{capt-of}

\usepackage{xpatch}
\usepackage{makecell}
\patchcmd{\thenomenclature}
  {\leftmargin\labelwidth}
  {\leftmargin\labelwidth\itemindent 20em }
  {}{}

  {\endMakeFramed}

\usepackage{nomencl}
\makenomenclature
\renewcommand{\nomgroup}[1]{%
    \ifthenelse{\equal{#1}{A}}{\item[\textbf{Greek Symbols}]}{%
    \ifthenelse{\equal{#1}{B}}{\item[\textbf{Other Variables}]}{}}} 
\RequirePackage{ifthen}

\makenomenclature
\makeatletter
\newcommand*{\rom}[1]{\expandafter\@slowromancap\romannumeral #1@}
\makeatother

\makeatletter
\DeclareRobustCommand{\textsupsub}[2]{{%
  \m@th\ensuremath{%
    ^{\mbox{\fontsize\sf@size\z@#1}}%
    _{\mbox{\fontsize\sf@size\z@#2}}%
  }%
}}
\makeatother

\begin{document}

\begin{frontmatter}



\title{From Attack to Protection: Leveraging Watermarking Attack Network for Advanced Add-on Watermarking}


% \author[label1,label2]{Seung-Hun Nam, Jihyeon Kang, Daesik Kim, Namhyuk Ahn, Wonhyuk Ahn}
\author[label1]{Seung-Hun Nam}
\author[label1]{Jihyeon Kang}
\author[label1]{Daesik Kim}
\author[label2]{Namhyuk Ahn\corref{cor1}}
\author[label1]{Wonhyuk Ahn\corref{cor1}}
\affiliation[label1]{organization={NAVER WEBTOON AI},
            city={Seongnam},
            country={South Korea}}

\affiliation[label2]{organization={Inha University},
            city={Incheon},
            country={South Korea}}

\cortext[cor1]{Corresponding authors}

% \author{} %% Author name

% \affiliation{organization={},%Department and Organization

%% Abstract
\begin{abstract}
Multi-bit watermarking (MW) has been designed to enhance resistance against watermarking attacks, such as signal processing operations and geometric distortions.
Various benchmark tools exist to assess this robustness through simulated attacks on watermarked images.
However, these tools often fail to capitalize on the unique attributes of the targeted MW and typically neglect the aspect of visual quality, a critical factor in practical applications.
To overcome these shortcomings, we introduce a watermarking attack network (WAN), a fully trainable watermarking benchmark tool designed to exploit vulnerabilities within MW systems and induce watermark bit inversions, significantly diminishing watermark extractability.
The proposed WAN employs an architecture based on residual dense blocks, which is adept at both local and global feature learning, thereby maintaining high visual quality while obstructing the extraction of embedded information.
Our empirical results demonstrate that the WAN effectively undermines various block-based MW systems while minimizing visual degradation caused by attacks. This is facilitated by our novel watermarking attack loss, which is specifically crafted to compromise these systems.
The WAN functions not only as a benchmarking tool but also as an add-on watermarking (AoW) mechanism, augmenting established universal watermarking schemes by enhancing robustness or imperceptibility without requiring detailed method context and adapting to dynamic watermarking requirements.
Extensive experimental results show that AoW complements the performance of the targeted MW system by independently enhancing both imperceptibility and robustness.
\end{abstract}


\end{frontmatter}



\section{Introduction}

Digital watermarking is a technique for protecting copyright by embedding identification information, known as watermark, into digital media, such as image, audio, and video \cite{wm_base1,wm_base2}.
Particularly, watermarking techniques targeting image have seen significant advancements.
This promising technique serves several purposes, primarily aimed at protecting copyright, verifying authenticity, and tracking transactions related to content purchases.
Image watermarking may manifest in two primary categories: visible watermarking and invisible watermarking.
Visible watermarking involves embedding a perceptible watermark into media, with logos and text appearing as overlays, thereby clearly signaling ownership to the viewer through the human visual system (HVS) \cite{kim2019robust, ahn2024imperceptible}.
In contrast, invisible watermarking is an approach that embeds imperceptible watermarks into media, typically designed to be undetectable by the HVS, ensuring that it does not cause perceptible alteration and affect the enjoyment of the original content \cite{wm_base3, zhang2024editguard}.

Invisible watermarking is categorized into zero-bit watermarking and multi-bit watermarking, based on the \emph{Watermark Capacity}, which denotes the amount of information encoded in the media.
In particular, multi-bit watermarking (MW), which inserts multiple watermark bits into media according to a predefined set of rules, has been actively studied due to its ability to extract multi-bit identification information from the watermarked image for the purpose of copyright protection \cite{wm_base4,li2024not,nam2018sift}.
In addition, MW embeds watermarks by considering the fundamental requirements: \emph{Imperceptibility}, which is the degree of invisibility of the watermarked signal, and \emph{Robustness}, which is the ability of the watermark to survive against various watermarking attacks \cite{wm_base3,kim2019robust,honig2024adversarial,pan2024finding}.

% BMVC 버전
% Digital watermarking is a technique used to protect copyright by embedding identification information, referred to as watermark, into the original image \cite{wm_base1,wm_base2}.
% Unlike visible watermarking, which inserts a watermark perceptible by the human visual system (HVS), invisible watermarking is an approach that embeds imperceptible watermarks \cite{kim2017blind}.
% In particular, multi-bit watermarking (MW), which is a representative example of invisible watermarking, has been actively researched so that multi-bit information can be extracted from the watermarked image \cite{wm_base4,nam2018sift}.
% MW inserts watermarks by considering the fundamental requirements: \emph{Imperceptibility}, which is the degree of invisibility of the watermarked signal, and \emph{Robustness}, which is the ability of the watermark to survive against various watermarking attacks \cite{wm_base3,kim2019robust}.

\begin{figure*}[t]
\centering{\includegraphics[width=0.98\linewidth]{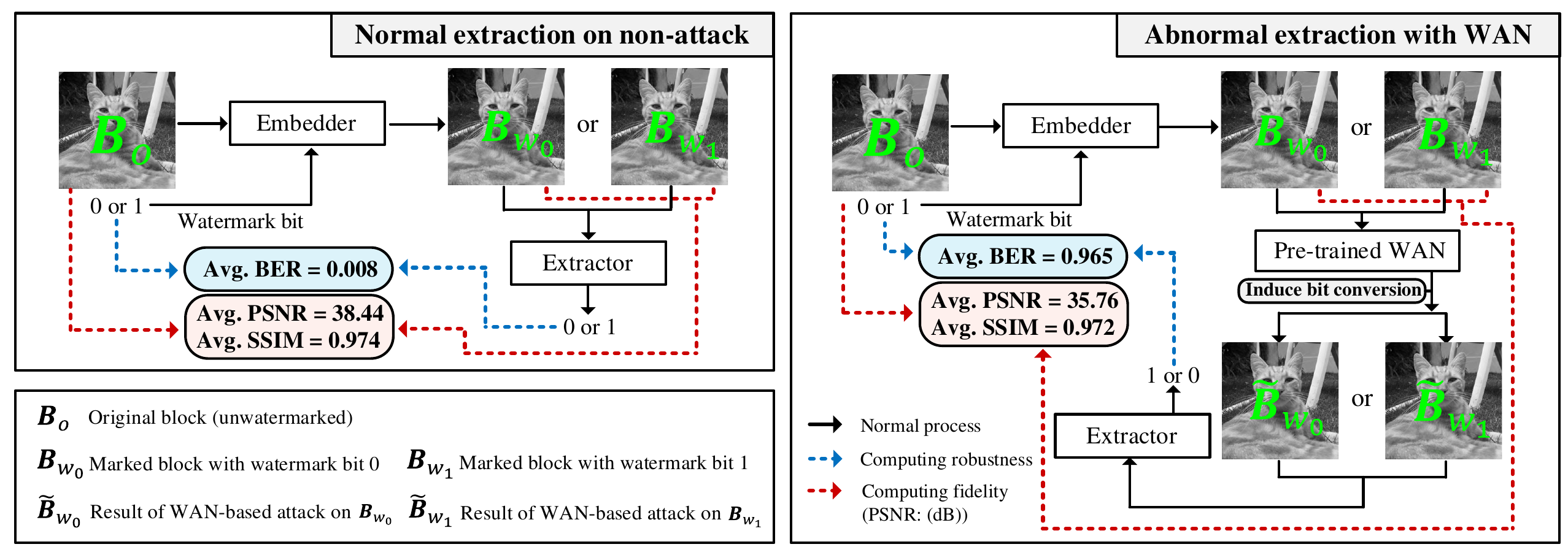}}
\caption{How the watermarking attack network (WAN) works. In contrast to the normal case where the embedded watermark bit is extracted correctly, the proposed WAN can enable attacks that induce abnormal extraction (e.g., 0$\rightarrow$1 or 1$\rightarrow$0) while minimizing visual degradation. Here, the case where the WAN is applied after a watermark bit is embedded into the original block $B_{o}$ is depicted.}
\label{figure_overview}
\end{figure*}

Imperceptibility of the watermarked image is assessed using image quality assessment (IQA) metrics (e.g., PSNR and SSIM \cite{ssim}), which appraise visual quality degradation caused by the watermark insertion.
In detail, the more negligible the discrepancies between the original and watermarked images, the higher the evaluation from the perspective of imperceptibility.
To assess robustness, benchmark tools composed of various watermarking attacks (e.g., signal processing operations and geometric distortions), such as StirMark \cite{stirmark1,stirmark2} and CheckMark \cite{checkmark}, are applied to watermarked images.
These tools assess the robustness of the watermarking system by evaluating the persistence of the watermark (i.e., whether the embedded watermarks can be normally extracted) under simulated attacks (see Figure~\ref{figure_overview}).
However, these tools attack watermarked images in a facile way without considering the context of the watermarking system, so they cannot dig into the specific weak points of the system \cite{attack_volo}.
From this, these attacks generally degrade the visual quality beyond what is acceptable for commercial usage in the process of interfering with watermark extraction by deteriorating the watermarked one.

\begin{figure}[t]
\centering{\includegraphics[width=1.0\linewidth]{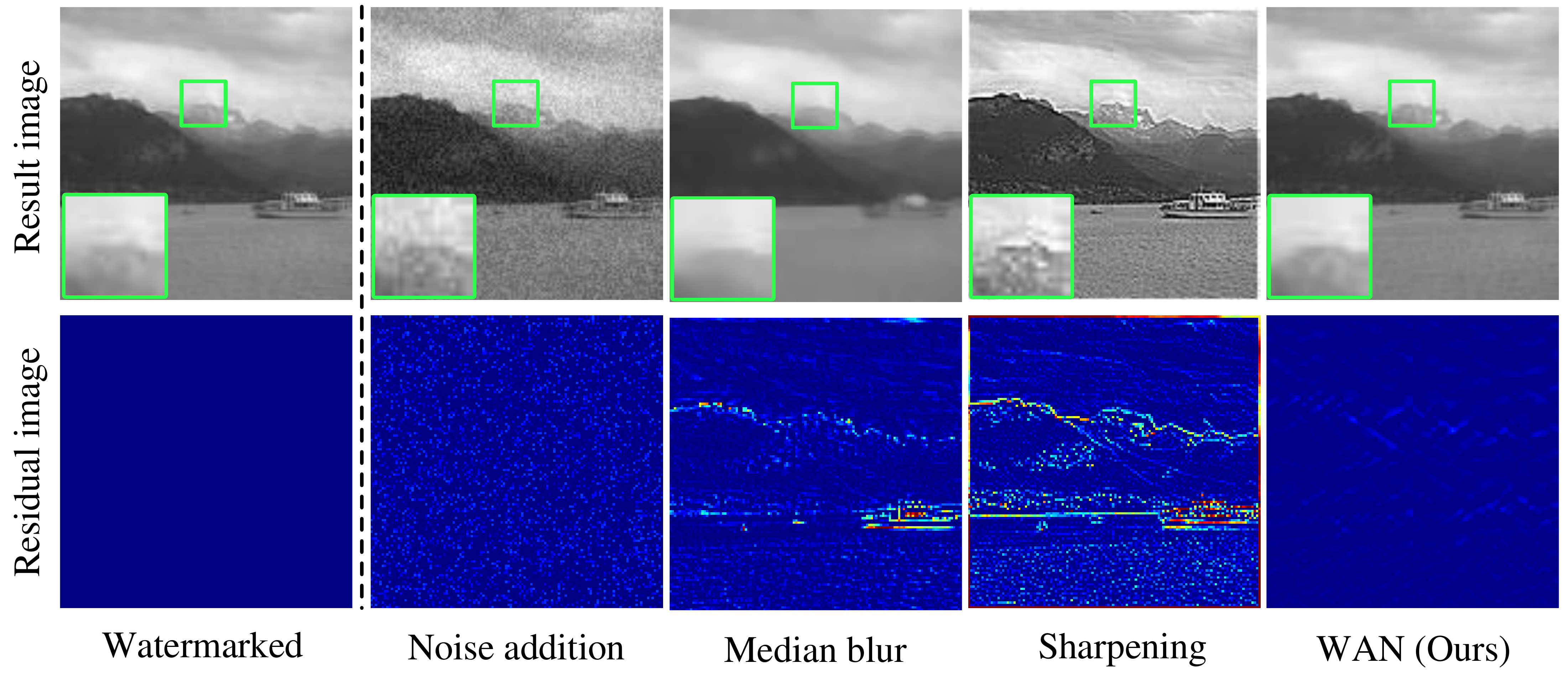}}
\caption{Visualization examples of conventional watermarking attacks and the proposed WAN-induced attack images. In this visual, residual images illustrate the differences between the watermarked image and its versions following various attacks.}
\label{figure_attack_example_intro}
\end{figure}

In light of the aforementioned case, malicious users (e.g., digital pirate) may try to design effective watermarking attack for interfering the watermark by targeting the MW while minimizing visual degradation \cite{attack_volo}, which further deepens the gap between attacks in the real-world and existing benchmark tools (see Figure~\ref{figure_attack_example_intro}).
To elaborate, malicious users may attempt to evade tracking by attacking the watermark embedded in pirated content, and in this process, they may try to devise sophisticated attacks that minimize visual degradation to maintain the quality necessary for unauthorized distribution.
To handle this, the designers of the watermarking system may assume a worst-case attack, where the watermark embedding and extraction algorithms are public (i.e., Kerckhoffs's principle \cite{kerckhoffs}), to make systems more robust against adversaries.
In this context, designing novel benchmark tools to create tests that are adequate for individual, specific watermarking systems to induce the false extraction of inserted information while maintaining a high quality level for the content is required.

\begin{table*}[t]
\caption{List of rule-based and CNN-based MW methods with corresponding attributes. As indicated in the table, this research verifies the performance of the proposed work across various MW systems, each characterized by distinct attributes. For CNN-based methods, watermark embedding is performed through a training phase, which is marked differently from traditional rule-based approaches.}
\centering
% \scriptsize
\resizebox{\linewidth}{!}{
\begin{tabular}{lccccl}\\
\Xhline{2\arrayrulewidth}
\multirow{2}{*}{\textbf{MW method}} &  \multirow{2}{*}{\textbf{Type}} & \textbf{Watermarking} & \textbf{Embedding} & \textbf{Extraction} & \textbf{Key}\\
 & & \textbf{domain} & \textbf{algorithm} & \textbf{approach} & \textbf{characteristic} \\
\hline
\textbf{M1} \cite{kim2018robust} & Rule & DCT & SS & Blind & Template matching\\
\textbf{M2} \cite{lin2011digital} & Rule & DCT & ISS & Blind & Multiple watermarks embedding \\
\textbf{M3} \cite{su2017improved} & Rule & QRD & DIF & Blind & Low false positive rate\\
\textbf{M4} \cite{makbol2016block} & Rule & DWT, SVD & DIF & Blind & Considering human visual system \\
\textbf{M5} \cite{nam2020nsct} & Rule & NSCT & QT & Blind & Perceptual masking \\
\textbf{M6} \cite{kim2012dtcwt} & Rule & DTCWT & QT & Blind & Robustness against geometric attacks \\
\textbf{M7} \cite{parah2016dct} & Rule & DCT & DIF & Blind & Robustness against hybrid attacks\\
\textbf{M8} \cite{wm_deep2} & CNN & \multicolumn{2}{c}{Attack simulator-based training}  & Blind & Adaptive robustness against attacks\\
% \textbf{M9} \cite{zhang2020udh} & CNN & \multicolumn{2}{c}{Attack layer-based end-to-end training}  & Blind & Versatility of data hiding \\
\Xhline{2\arrayrulewidth}
\end{tabular}
}
\label{table1}
\end{table*}

Motivated by the need for watermarking designers, who want to analyze the vulnerability of their MW methods, we propose a watermarking attack network (WAN) that exploits the weak points of individual watermarking systems.
As illustrated in Figure~\ref{figure_overview}, the proposed WAN is devised to hinder the extraction of inserted watermarks by adding interference signals to mislead the watermarking extractor without compromising visual quality.
With proposed loss function, which consists of both watermarking attack loss and content loss, our work can both induce abnormal extraction (i.e., inducing the extraction of the inserted watermark bits by inverting them) and generate a reconstructed image with a visual quality similar to the original content.
We determined that the residual dense block-based architecture’s ability \cite{rdn} to learn local and global features is suitable for analyzing the characteristics of each MW method, which is composed of various procedures and detailed attributes, such as the watermarking domains and embedding algorithms  (as detailed in Table~\ref{table1}).
The proposed WAN is well-suited as a benchmark tool for performance evaluation in the watermarking design process, as it enables sophisticated attacks on MW methods with various attributes\textemdash specifically, by obstructing watermark extraction while maintaining visual quality.

% Moreover, we show that our WAN can be used as an add-on module to further improve the performance of existing MW methods, which we designate as Add-on Watermarking (AoW).
Furthermore, we introduce Add-on Watermarking (AoW), which enhances the performance of existing watermarking systems through the support of the WAN framework.
It offers users a dynamic choice between robustness and invisibility, depending on their specific requirements.
Based on the residual data with original, watermarked, and attacked images, AoW can achieve improvements in terms of robustness or imperceptibility of the predefined MW methods without modifying the algorithm, such as adjusting specific parameters or watermarking processes.
Based on WAN's attack capabilities, AoW selectively enhances the performance of watermarking techniques, embodying the paradigm shift \textit{from exploiting vulnerabilities to reinforcing protection}.
% Based on WAN's attack capabilities, AoW selectively enhances the performance of watermarking techniques, embodying the concept of \textit{from attack to protection}.
Particularly, rule-based MW techniques generally secure invisibility and robustness by experimentally exploring and fixing various processes and parameters in advance; however, the proposed AoW enables performance improvement without altering these established rules.

This work is an extended version of our previous work \cite{nam21wan} with improvements made in the following several ways:
\begin{itemize}
% \item More detailed analyses are presented in the ``Background section'', which now include a concise summary of rule-based and CNN-based MW techniques, an analysis of traditional watermarking attacks included in benchmark tools, and a review of the latest literature.
\item The extended paper now includes a comprehensive investigation, encompassing a concise summary of rule-based and CNN-based MW techniques, a detailed examination of traditional watermarking attacks featured in existing benchmark tools, and a thorough review of the latest literature.
\item A new section was introduced to present the concept and methodology of AoW. This section provides a detailed introduction to the methodologies for enhancing the invisibility and robustness of MW systems through the use of WAN in AoW.
% \item Experiments on the conventional rule-based methods \textbf{M6} \cite{kim2012dtcwt} and \textbf{M7} \cite{parah2016dct}, which possess attributes not considered in the previous version, have been added. To further demonstrate the generalization performance of our WAN, the experiments on convolutional neural network (CNN)-based MW systems \textbf{M8} \cite{wm_deep2} and \textbf{M8} \cite{zhang2020udh} have been conducted.
\item A comprehensive analysis was conducted using various MW systems to emphasize the effectiveness of the proposed WAN and AoW. In particular, To further demonstrate the generalization performance of our work, experiments on rule-based and convolutional neural network (CNN)-based MW systems (i.e., \textbf{M6} -- \textbf{M8}), which were not considered in the previous version, have been conducted.
% \item To compare with recently proposed watermarking attacks, we modified the our existing approach of inducing watermark bit inversion (e.g., BER=1) and designed a loss function that causes abnormal extraction at the level of random guessing (e.g., BER=0.5). 
\end{itemize}

The main contributions are listed as follows.
\begin{itemize}
\item We propose a CNN-based watermarking attack framework designed to interfere the extraction processes of MW systems. This framework introduces an innovative methodological approach for probing and enhancing the security paradigms of digital watermarking schemes, highlighting the potential vulnerabilities against neural network-driven attacks.
% To the best of our knowledge, this is the first attempt to successfully introduce a convolutional neural network (CNN)-based watermarking attack framework for interfering with the watermark extraction of MW systems.
\item The proposed WAN induces abnormal watermark extraction while conserving perceptual quality, compared to widely utilized existing benchmark tools \cite{stirmark1,stirmark2}. 
For specific MW methods, particularly those based on rule-based and AI-driven approaches, the WAN can apply subtle modification to induce the extraction of the watermark bit embedded in the image in an inverted state (e.g., 0$\rightarrow$1 or 1$\rightarrow$0).
\item WAN serves as an add-on to watermarking schemes, enhancing their performance and embodying the concept of \textit{from attack to protection}. Pre-trained WAN can be used as an add-on module, known as AoW, which yields performance gains in terms of imperceptibility and robustness for the targeted rule-based MW method.
% \item We present one possible usage of WAN: a pre-trained WAN can be used as an add-on module, namely AoW, which yields performance gains in terms of imperceptibility and robustness in existing rule-based MW methods.
\item To ensure rigorous experiments, an extensive dataset was constructed based on various watermarking techniques, and experiments (e.g., quantitative and qualitative evaluations of WAN, results on color images, results on 16 bits capacity, ablation study, and results of AoW) conducted on this dataset demonstrated the superiority of our work.
\end{itemize}

The remainder of this paper is organized as follows.
Section~\ref{sec_background} presents background relevant our work.
The proposed watermarking attack framework and objective function are presented in Section~\ref{sec_proposed_method}.
Our add-on watermarking framework, which utilizes the proposed WAN to enhance imperceptibility or robustness, is described in Section~\ref{sec_aow}.
The performance of our work is demonstrated in Section~\ref{sec_experiments}.
Finally, Section~\ref{sec_Conclusion} concludes this paper.

\begin{figure*}[t]
\centering{\includegraphics[width=0.98\linewidth]{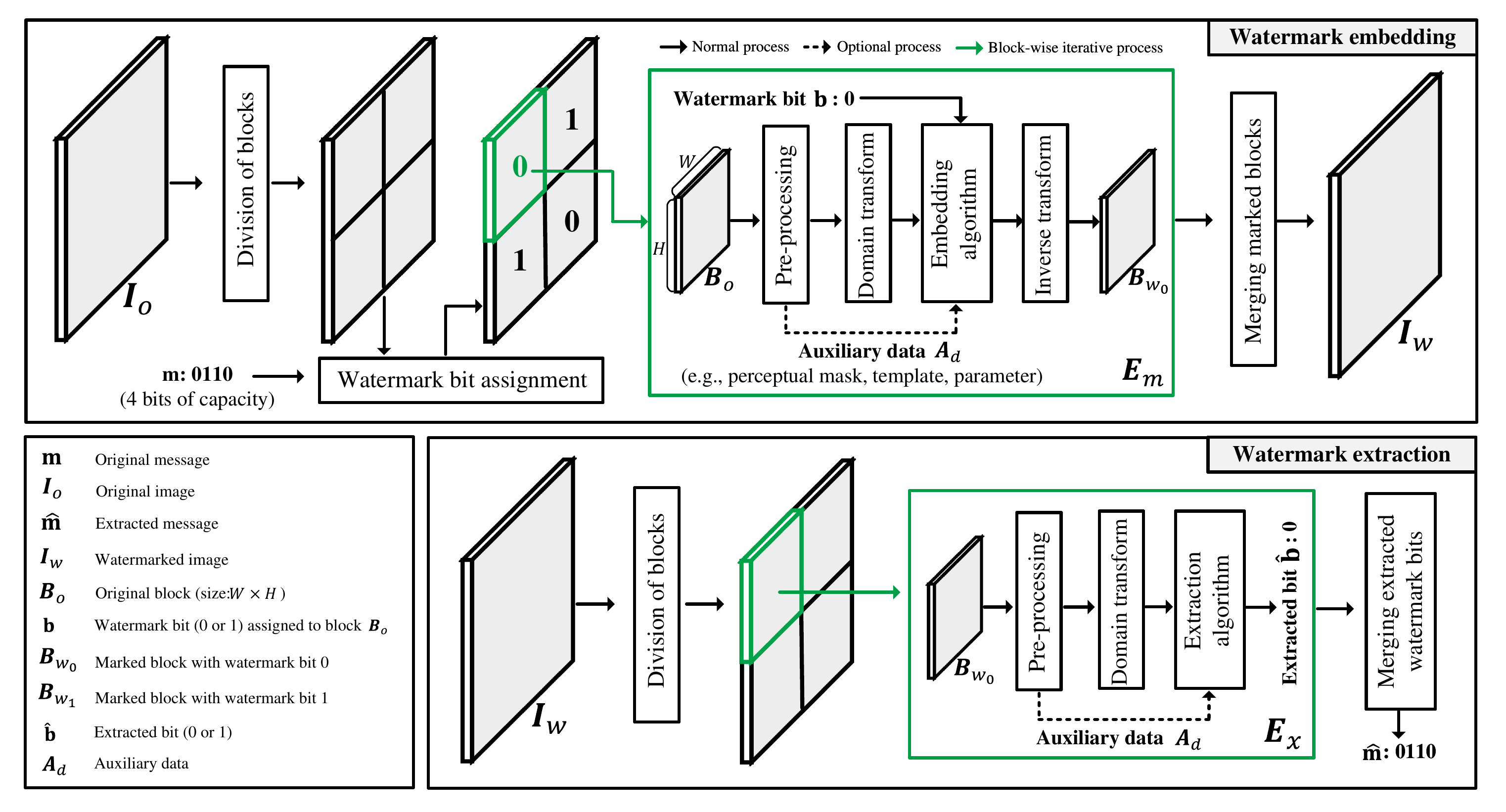}}
\caption{A general overview of the conventional rule-based MW system, where watermark bits are inserted and extracted in a block-wise manner.}
\label{figure_multibit_watermarking}
\end{figure*}

\section{Background}
\label{sec_background}
In this section, we review the concept of rule-based and CNN-based MW approaches and the limitations of existing benchmark tools for testing robustness against watermarking attacks.

\vspace{-2mm}

\subsection{Multi-bit Watermarking}

\subsubsection{Rule-based approach}

Rather than using zero-bit watermarking to detect the presence or the absence of a watermark, MW can be used in various applications since the n-bit-long message ($\mathbf{m}=\{0,1\}^n$) can be inserted in the host image  $I_{o}$ to get a watermarked image $I_{w}$ (see Figure~\ref{figure_multibit_watermarking}).
In particular, block-based MW methods \cite{lin2011digital,kim2018robust,makbol2016block,su2017improved,nam2020nsct}, which insert a watermark bit (0 or 1) in each original block $B_{o}$, are mainly used for multi-bit information insertion rather than the keypoint-based approach \cite{nam2018sift} due to the benefits that can be achieved by utilizing the entire domain.
As shown in Table~\ref{table1}, the attributes of block-based MW methods (from \textbf{M1} to \textbf{M7}) vary, and each attribute is determined by considering the aimed performance and media's inherent properties.
In general, the transform domain such as discrete wavelet transform (DWT) \cite{shensa1992discrete}, discrete cosine transform (DCT) \cite{ahmed1974discrete}, nonsubsampled contourlet transform (NSCT) \cite{nsct,karim2024infrared}, dual-tree complex wavelet transform (DT-CWT)~\cite{selesnick2005dual}, singular value decomposition (SVD) \cite{svd}, and QR decomposition (QRD) \cite{de1992generalizations} is first applied to each pre-processed block, and then watermark embedding is performed by applying an embedding algorithm such as spread spectrum (SS) \cite{wm_base3,wm_base2}, improved spread spectrum (ISS) \cite{iss}, quantization (QT) \cite{qt1,qt2}, and embedding for causing differences between sub-groups (DIF) \cite{su2017improved,makbol2016block} to the selected domain.

The block-based MW system consists of an embedder and extractor, and procedures for watermark embedding and extraction are performed independently for each block (see Figure~\ref{figure_multibit_watermarking}).
The block-wise embedding function follows: $B_w = E_m(B_o,\mathbf{b},A_d)$ where $\mathbf{b}$ and $A_d$ denote an assigned watermark bit and auxiliary data, respectively.
By applying $E_m$ to each $B_o$, constituting $I_o$, $I_w$ containing message $\mathbf{m}$ with a capacity of $n$ can be obtained.
In the case where $n$ is set to 1, $I_w$ and $B_w$ are equal.
In the extraction phase, the estimated message $\mathbf{\hat{m}}$ can be obtained by merging $\mathbf{\hat{b}}$, which is extracted from each $B_w$, and the block-wise extraction function takes the following form in the case of a blind fashion: $\mathbf{\hat{b}} = E_x(B_w, A_d)$ where the blind fashion denotes that the original image is not required \cite{wm_base6}.
In the case where $n$ is set to 1, $\mathbf{\hat{m}}$ and $\mathbf{\hat{b}}$ are equal.
The performance of MW is evaluated in terms of imperceptibility and robustness.
Specifically, the visual differences between the original and watermarked images are determined using the IQA metrics, such as peak signal-to-noise ratio (PSNR) and structural similarity (SSIM) \cite{ssim}, and robustness is evaluated by calculating the bit error rate (BER) between $\mathbf{m}$ and $\mathbf{\hat{m}}$.

\begin{figure*}[t]
\centering{\includegraphics[width=0.99\linewidth]{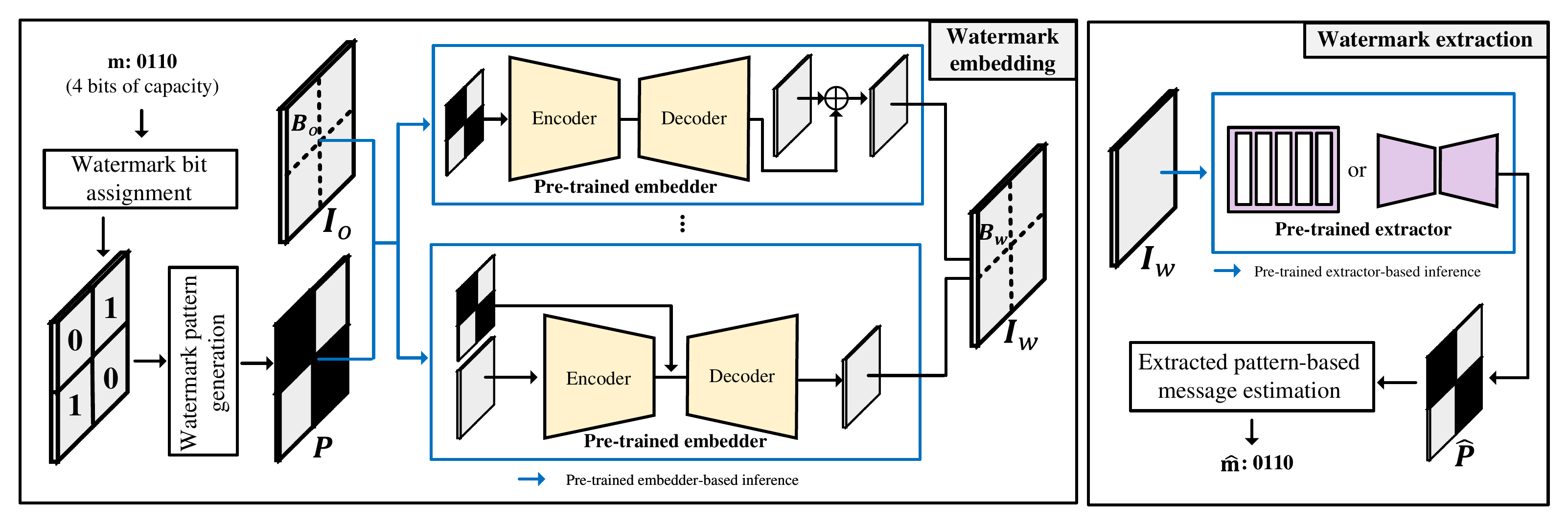}}
\caption{A general overview of the neural network-based MW system, where watermark bits are assigned to block-wise watermark pattern images. Here, $P$ and $\hat{P}$ denote original watermark pattern and extracted watermark pattern, respectively.}
\label{figure_multibit_watermarking_CNN}
\end{figure*}

\subsubsection{Neural Network-based approach}

With the advances of data-driven learning methodologies, neural network-based MW methods \cite{wm_deep1,wm_deep2,zhang2020udh,chen2024secure} have emerged as a promising solution, offering significant improvements in both invisibility and robustness.
Particularly, these approaches focus on designing network architectures such as embedders and extractors (see Figure~\ref{figure_multibit_watermarking_CNN}), and guiding the models through training to achieve high imperceptibility and robustness against various attacks.
Imperceptibility is commonly achieved by designing a perceptual loss that minimizes the visual differences between the original and watermarked images, while robustness against attacks is significantly ensured through staged training based on an attack simulator (i.e., approach that utilize real-world watermarking attacks, not in the form of network layers) or differentiable attack layer-based end-to-end training \cite{wm_deep2,zhang2020udh,zhu2018hidden}.
Specifically, to ensure robustness against attacks, an objective function that reduces the differences between the original watermark pattern $P$ and the extracted pattern $\hat{P}$ estimated from the attacked watermarked image is devised \footnote{This study focuses on analyzing the performance of the proposed WAN and AoW for rule-based MW watermarking that operates on a block basis, adopting a widely utilized method that generates a grid-shaped watermark pattern $P$ for a given message $m$.}.
Upon completion of the training of the embedder and extractor networks, which utilize an extensive training dataset containing original images and watermark patterns (considering combinations of $b$ that compose $m$), watermark embedding and extraction are completed by feeding the necessary inputs into the embedder and extractor, as specified in Figure~\ref{figure_multibit_watermarking_CNN}, to perform model inference.
The performance of neural network-based MW can be evaluated using IQA metrics and BER, similar to rule-based MW, and as specified in the figure, the processes of generating the watermark pattern $P$ for $m$ and estimating $\hat{m}$ through the extracted $\hat{P}$ adhere to predefined rules.

\begin{figure*}[t]
\centering{\includegraphics[width=0.98\linewidth]{./figure_v2/figure_visual_degradation_stirmark.pdf}}
\caption{Comparison of fidelity between StirMark attacks causing extraction of random guessing and WAN. Here, the watermarking method employed is \textbf{M2}.}
\label{figure_visual_degradation_stirmark}
\end{figure*}

\subsection{Watermarking Attack and Motivation}
\label{sec_watermarking_attack_motivation}
Watermarking attacks are employed to evaluate the robustness of MW methods; let $\tilde{B}_w$ be the attacked block image of $B_w$.
By comparing the watermark bit extracted from $B_w$ and $\tilde{B}_w$, a MW designer can evaluate the robustness of the MW by determining whether the hidden information survived \cite{attack_volo,wang2024spy}.
Currently, StirMark \cite{stirmark1,stirmark2} and CheckMark \cite{checkmark} are the representative benchmark tools that provide various types of common attacks such as signal processing operations and geometric distortions.
As can be seen in Figure~\ref{figure_visual_degradation_stirmark}, common watermarking attacks mounted in StirMark are accompanied by visual degradation and have a limitation of not being able to model the vulnerabilities of each MW method.
That is, the more that a watermarking attack utilizes the characteristics of the targeted watermarking system, the more effective the attack is possible without image quality degradations.

% \begin{figure*}[t]
% \centering{\includegraphics[width=0.75\linewidth]{./figure_v2/figure_multibit_watermarking_CNN_based.png}}
% \vspace{-3mm}
% \caption{\color{red}Motivation and concept of WAN + AoW\color{red}}
% \vspace{-3mm}
% \label{figure_motivation}
% \end{figure*}

% Recent studies have been presented on watermarking attacks from the perspective of reducing perceptual quality degradation.
With the development of neural networks, CNN-based MW methods \cite{wm_deep1,wm_deep2,zhang2020udh} have been newly proposed, and they can be neutralised with adversarial attacks attempting to fool watermarking systems through malicious inputs; these are referred to as adversarial examples. 
However, attacking numerous handcrafted MW methods that contain non-differentiable operators with an adversarial attack is difficult \cite{nam21wan}.
% For example, many factors make existing traditional MW hard to be trainable with back-propagation.
% First, many MWs contain quantization as a key operation when embedding, which is non-differentiable.
% Second, many MWs contain irregular repetitive operations which means gradients easily vanish.
% Third, many MWs including SVD and the derivatives of the eigenvectors tend to be numerically unstable.
For example, several factors make existing traditional MW hard to be trainable with back-propagation.
Many MWs involve quantization as a crucial embedding operation, which is inherently non-differentiable.
Additionally, these methods often contain irregular repetitive operations, leading to easily vanishing gradients.
Furthermore, many MWs including SVD and the derivatives of the eigenvectors tend to be numerically unstable.

Considering these factors, concepts for deep learning-based attacks applicable to handcrafted MW have recently been researched.
A differential evolution-based attack, which randomly modifies one pixel and queries the extractor, has been proposed \cite{su2019one}.
However, this approach faces limitations in disabling robust MW systems with only minimal pixel alterations.
Focusing on watermarking techniques that inject noise-like signals into original images to embed identification information, researchers have proposed attack approaches aimed at aligning the distribution of watermarked images more closely with that of original images.
CNN-based approaches are detailed in \cite{hatoum2021using,geng2020real}, and a GAN-based method in \cite{li2021concealed}.
These attacks enable the removal of watermarks, thereby producing resultant images that mimic the original distribution and can lead to abnormal extraction, with a BER equivalent to random guessing at 0.5.
However, these have a limitation in that they cannot facilitate advanced attacks capable of inverting the embedded information, which malicious users might execute to disrupt content distributor.

In this study, to address the issues mentioned, we propose a novel watermarking attack concept, \textbf{WAN}, and a concept leveraging the performance of pre-trained WAN, \textbf{AoW}, to enhance the performance of the target MW system.
The distinct advantages of these approaches are outlined as follows.

\begin{itemize}
\item Our WAN considers a more difficult environment than that of traditional attack simulators (i.e., StirMark), where malicious users try to interfere watermarks without degradation of visual quality.
In particular, rather than guiding attacked images to mimic the characteristics of neutral content, our work induces the inversion of the embedded watermark bits to facilitate abnormal extraction; hence, it can offer watermarking framework designers a comprehensive benchmark tool.
\item With the aid of WAN, which analyzes and intentionally interferes with MW composed of various attributes and sophisticated processes, our AoW facilitates performance enhancements in the imperceptibility and robustness of the target MW system.
Particularly in the case of rule-based MW, where composite processes and parameters are optimized empirically, modifying rules to enhance performance entails significant resource investment. However, the proposed AoW can yield performance gains without altering these established rules.
\end{itemize}

% \begin{framed}[.42\textwidth]
% \vspace{-3.5mm}
% \nomenclature[01]{$I_{o}$}{Original image}
% \nomenclature[02]{$I_{w}$}{Watermarked image}
\nomenclature[03]{$B_{o}$}{Original block ($W\times H$)}
\nomenclature[04]{$B_{w_0}$}{Watermarked block with bit 0}
\nomenclature[05]{$B_{w_1}$}{Watermarked block with bit 1}
\nomenclature[06]{$\tilde{B}_{w_0}$}{Result of WAN-based attack on $B_{w_0}$}
\nomenclature[07]{$\tilde{B}_{w_1}$}{Result of WAN-based attack on $B_{w_1}$}
\nomenclature[08]{$R_{o,w_0}$}{Residual between $B_{o}$ and $B_{w_0}$}
\nomenclature[09]{$R_{o,w_1}$}{Residual between $B_{o}$ and $B_{w_1}$}
\nomenclature[10]{$\tilde{R}_{o,w_0}$}{Residual between $B_{o}$ and $\tilde{B}_{w_0}$}
\nomenclature[11]{$\tilde{R}_{o,w_1}$}{Residual between $B_{o}$ and $\tilde{B}_{w_1}$}
\printnomenclature[1.2cm]
% \end{framed}

\begin{figure*}[t]
\centering{\includegraphics[width=0.99\linewidth]{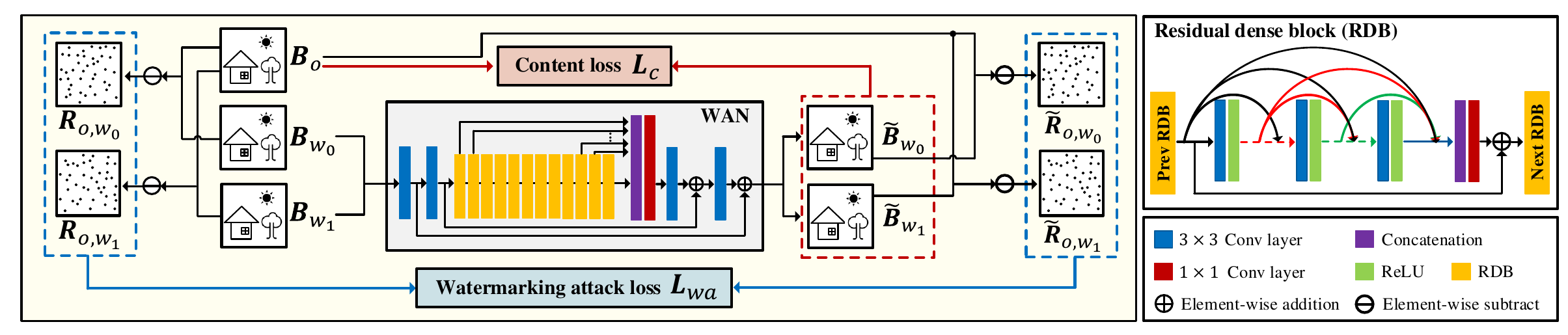}}
\caption{Schematic illustration of the training WAN framework.}
\label{figure_WAN_framework}
\end{figure*}

\section{Watermarking Attack Network (WAN)}
\label{sec_proposed_method}
The proposed WAN targets block-based MW and needs one triple set of block images, of $B_o$, watermarked images with bit 0 $B_{w_0}$, and watermarked images with bit 1 $B_{w_1}$, in the training phase.
WAN takes $B_{w_0}$ and $B_{w_1}$ as inputs and reconstructs each of them into attacked images $\tilde{B}_{w_0}$ and $\tilde{B}_{w_1}$, respectively.
Our goal is to reconstruct images that mislead the watermarking extractor to decide on the wrong bit.
In other words, when $\tilde{B}_{w_0}$ and $\tilde{B}_{w_1}$ are considered to have been inserted 1 bit and 0 bit, we judge the attack to be successfully done.
On the other hand, the attacked image should be similar to the original to minimize visual degradation.

Figure~\ref{figure_WAN_framework} illustrates the training WAN framework, which includes a mini-batch of triple set consisting of $B_o$, $B_{w_0}$, and $B_{w_1}$, RDN-based WAN, and the proposed loss functions.
From the viewpoint of RDN-based WAN only, the input is just a watermarked image (e.g., $B_{w_0}$), and it is expected to produce an attacked image ($\tilde{B_{w0}}$) with an inverted bit at the watermark extractor.
The lines outside RDN-based WAN are drawn to pair for each bit 0 and 1 cases for loss function in the triple set configuration.
In this section, we start with in-depth descriptions of loss functions consisting of watermarking attack loss and content loss and provide detailed descriptions of the architecture of the network and the mini-batch configuration.

\subsection{Loss Function}
\label{sec_loss}

The proposed WAN is trained to reconstruct attacked images containing an inverted watermark bit while minimizing the visual quality degradation.
To achieve this, we propose a customized loss as an objective function to train the WAN as follows:
\begin{equation}
\label{eq0}
\mathcal{L} = \lambda_{wa}\mathcal{L}_{wa} + \lambda_{c}\mathcal{L}_{c},
\end{equation}
where $\mathcal{L}_{wa}$ and $\mathcal{L}_{c}$ represent watermarking attack loss, which is devised to change an inserted bit and content loss to minimize visual degradation, respectively.
$\lambda_{wa}$ and $\lambda_{c}$ indicate predefined weight terms for each loss.

\subsubsection{Watermarking Attack Loss}
\label{sec_watermarking_attack_loss}

Existing watermarking methods vary in terms of the watermarking domains and embedding algorithms, so it is difficult to theoretically model MW in a single system.
Moreover, conventional MW methods incorporate non-differentiable operations, so it is difficult for the neural network to learn directly from these methods even though step-by-step instructions are publicly available.
We simplify this problem as the watermark signal is added to the original image in the pixel domain, and focus on the noise patterns that are decided by bit information.
In other words, the residual signal arose by bit 0 insertion $R_{o,w_{0}}= B_o-B_{w_0}$ and the residual signal arisen by bit 1 insertion $R_{o,w_{1}}= B_o-B_{w_1}$, which can be identified by neural networks.
We hypothesise that the neural network can remove watermark signals in images and insert opposite noise patterns, which causes wrong bit extraction at the watermarking extractor.
In this case, the attacked image $\tilde{B}_{w_0}$ on $B_{w_0}$ would have similar noise pattern $\tilde{R}_{o,w_{0}}=B_o-\tilde{B}_{w_0}$ to $R_{o,w_{1}}$, for which the one with bit 0 makes.
The noise pattern $\tilde{R}_{o,w_{1}}$ of the attacked image $\tilde{B}_{w_1}$ on $B_{w_1}$ would be similar to $R_{o,w_{0}}$, in the same way.

To capture the above observation, watermarking attack loss for the image of size $W\times H$, $\mathcal{L}_{wa}$ is defined as follows:
\begin{equation}
\label{eq1}
\mathcal{L}_{wa} = \frac{1}{N}\sum_{i=1}^{N}|R^{i}_{o,w_{0}}-\tilde{R}^{i}_{o,w_{1}}| + \frac{1}{N}\sum_{i=1}^{N}|R^{i}_{o,w_{1}}-\tilde{R}^{i}_{o,w_{0}}|,
\end{equation}
where superscript $i$ refers to pixel location and $N=W\times H$.
The first term of the equation is for deriving the watermark bit 1 inserted in $B_{w_1}$ into 0, and the second term is for deriving bit 0 inserted in $B_{w_0}$ into bit 1.
As depicted in Figure~\ref{figure_WAN_framework}, a loss is designed by pairing the residual images before and after going through the WAN according to the inserted bit and reducing the difference between the paired images.
Through the $\mathcal{L}_{wa}$, it is possible to add a fine noise-like attack that inverts the actually inserted bit during the process of passing the watermarked images over the WAN.

In detail, when the watermarking attack loss is changed from residual differences into pixel differences, we found that the speed of convergence gets slower and that the effectiveness of watermarking attack degrades (see Section~\ref{residual_data}).
This is because the scale and variance of outputs with pixel differences become bigger than the ones with residual differences, which makes WAN training more difficult.
We chose residual differences for watermarking attack loss, and this study was performed based on signed residual images.

\subsubsection{Content Loss}

In terms of watermarking attack, it is important to preserve visual quality while adding adversarial signals.
To this end, content loss is adopted to reduce the visual differences between the original content $B_{o}$ and its corresponding reconstructed images, including $\tilde{B}_{w_0}$ and $\tilde{B}_{w_1}$ attacked by the WAN (see Figure~\ref{figure_WAN_framework}).
Inspired by the papers in \cite{l1_loss} demonstrating that $\ell_{1}$ loss can bring better visual quality than $\ell_{2}$ loss for general restoration tasks, the content loss of $\mathcal{L}_{c}$ is defined as follows:
\begin{equation}
\label{eq_c}
\mathcal{L}_{c} = \frac{1}{N}\sum_{i=1}^{N}\sum_{j=0}^{1}|B^{i}_{o}-\tilde{B}^{i}_{w_{j}}|.
\end{equation}
From $\mathcal{L}_{c}$, it is possible to conduct a watermarking attack while minimizing visual quality degradations in the original content.
Through the final objective function of $\mathcal{L}$ combined with $\mathcal{L}_{c}$ and $\mathcal{L}_{wa}$, the proposed WAN can reconstruct images in a way that adversely affects the extraction of the inserted bit while maintaining the inherent properties of the original content.

\subsection{Model Architecture}
\label{sec_architecture}
We follow the network design from the residual dense network (RDN) that is used for the learning of the local and global features and the ability of image restoration \cite{rdn,zhang2020residual}.
The residual dense block (RDB) constituting the RDN is composed of densely connected convolution (Conv) layers and is specialized in extracting abundant local features.
In the proposed WAN, the pooling layer and up-sampling are excluded, so the input and output sizes are the same $(\{B_{w_0},B_{w_1},\tilde{B}_{w_0},\tilde{B}_{w_1}\}$ $\in \mathbb{Z}^{1\times W\times H})$.
The first and second Conv layers are placed to extract shallow features and conduct global residual learning.
Next, RDBs are placed consecutively, and we expect sub-components for local residual learning and local feature fusion commonly used in RDB to help our model learn fine-grained distortions caused by watermark embedding.
After that, by the concatenation layer followed by $1\times1$ and $3\times3$ Conv layers, dense local features extracted from the set of RDBs are fused in a global way.
The deep part of the proposed WAN is composed for global residual learning based on shallow feature maps.

\subsection{Mini-batch Configuration}

Since invisible MW is the approach of inserting a watermark so that it is unnoticeable by HVS, mini-batch configuration suitable for fine signal learning is required instead of the standard mini-batch used in high-level computer vision.
The authors in \cite{paired} presented paired mini-batch training, which is efficient for learning noise-like signals such as multimedia forensics \cite{bae2021dual,yoon2021frame} and steganalysis \cite{yu2020bitmix}.
To aid in learning the discriminative features between watermarked results more effectively, paired mini-batch training is employed.
That is, $B_{w_0}$ and $B_{w_1}$ generated for the same original image $B_{o}$ are allocated in a single batch, which allows the proposed WAN to learn fine signals due to the differences in the fine signals caused by the watermark bit.
In detail, when the batch size is $b_{s}$, $\frac{b_{s}}{2}$ $B_{w_0}$ images are selected first, and then $\frac{b_{s}}{2}$ $B_{w_1}$ images corresponding to $B_{w_0}$ are assigned to be in the same batch.
The entire dataset is shuffled every epoch.

\begin{figure*}[t]
\centering{\includegraphics[width=0.98\linewidth]{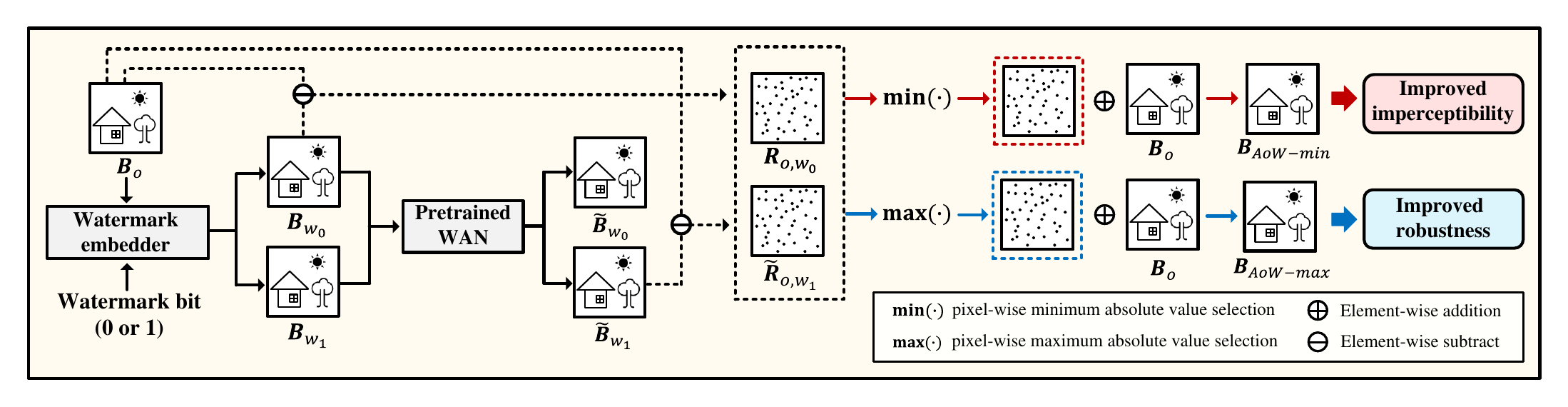}}
\vspace{-3mm}
\caption{Schematic illustration of the proposed AoW.}
\label{figure_AoW}
\end{figure*}

\section{Add-on Watermarking (AoW)}
\label{sec_aow}

As described in subsection~\ref{sec_watermarking_attack_motivation}, for rule-based MW methods, achieving further performance enhancements requires experimental exploration of complex processes and parameters, making it challenging to alter predefined rules.
To elaborate further, watermarking designers need to adjust their watermarking methods according to the requirements of content characteristics and distribution environments.
Professional photographers may prefer the invisibility of embedded watermarks for robustness against various signal processing attacks. 
On the other hand, confidential documents have to be very robust against not only general processing but also malicious editing, so visual degradation to some extent may be acceptable.

Considering these needs, we present Add-on Watermarking (AoW), one application of WAN, that adjusts between the imperceptibility and robustness of pre-defined watermarking methods.
With the support of WAN, which analyzes and deliberately interferes with MW systems comprising diverse attributes and intricate processes, our AoW effectively enhances the performance of the target MW system.
As illustrated in Figure~\ref{figure_AoW}, additional performance gain is carried out by adding noise-like residual information to the original blocks, resulting in an efficient process that does not necessitate additional rule modifications or training for the watermark embedder.

% We introduce AoW-min and AoW-max, which are defined as follows:
We introduce AoW-min and AoW-max for specific enhancements: AoW-min to improve imperceptibility, and AoW-max to enhance robustness.
The results of these two add-on modules are represented by $B_{AoW-min}$ and $B_{AoW-max}$, respectively, and are defined as follows.

\begin{algorithm}[t]
\caption{Performance enhancement process of the proposed AoW}
\textbf{Input:} Original block $B_{o}$, Watermark bit $b$ (0 or 1), Auxiliary data $A_d$\\
\textbf{Output:} Result of AoW-min $B_{AoW-\text{min}}$, Result of AoW-max $B_{AoW-\text{max}}$

\begin{algorithmic}[1]
\STATE $B_{w_o}, B_{w_1} \gets E_{m}(B_o,b,A_d)$ \color{gray}$E_{m}:$ watermark embedder\color{black}
\STATE $\tilde{B}_{w_0}, \tilde{B}_{w_1} \gets \text{WAN}(B_{w})$ \color{gray}WAN: pre-trained WAN-based model inference, $B_{w}$ involves $B_{w_o}, B_{w_1}$\color{black} 
\STATE $R_{o,w_0}, R_{o,w_1}, \tilde{R}_{o,w_0}, \tilde{R}_{o,w_1} \gets$\\
$\text{RES}(B_o, B_{w_o}, B_{w_1}, \tilde{B}_{w_0}, \tilde{B}_{w_1})$ \color{gray}RES: function to calculate residual data\color{black}
\vspace{0.8mm}
\FOR{$k = 0$ to $1$}
\FOR{$i = 0$ to $W-1$}
\FOR{$j = 0$ to $H-1$}
% \IF{$|R^{i,j}_{o,w_{k}}| < |\tilde{R}^{i,j}_{o,w_{1-k}}|$}
% \STATE $B^{i,j,k}_{AoW-min} \gets B^{i,j}_{o} + R^{i,j}_{o,w_{k}}$
% \ELSE
% \STATE $B^{i,j,k}_{AoW-min} \gets B^{i,j}_{o} + \tilde{R}^{i,j}_{o,w_{1-k}}$
% \ENDIF
% \IF{$|R^{i,j}_{o,w_{k}}| < |\tilde{R}^{i,j}_{o,w_{1-k}}|$}
% \STATE $B^{i,j,k}_{AoW-max} \gets B^{i,j}_{o} + \tilde{R}^{i,j}_{o,w_{1-k}}$
% \ELSE
% \STATE $B^{i,j,k}_{AoW-max} \gets B^{i,j}_{o} + R^{i,j}_{o,w_{k}}$
% \ENDIF
\STATE $B^{i,j,k}_{AoW-min} \gets B^{i,j}_{o} + \min(|R^{i,j}_{o,w_{k}}|, |\tilde{R}^{i,j}_{o,w_{1-k}}|)$
\vspace{-0.4mm}
\STATE $B^{i,j,k}_{AoW-max} \gets B^{i,j}_{o} + \max(|R^{i,j}_{o,w_{k}}|, |\tilde{R}^{i,j}_{o,w_{1-k}}|)$
\ENDFOR
\ENDFOR
\ENDFOR
\STATE \textbf{return} $B_{AoW-\text{min}}$, $B_{AoW-\text{max}}$
\end{algorithmic}
\label{algorithm1}
\end{algorithm}

\begin{equation}
\label{eq_aow_min}
B^{i,j,k}_{AoW-min} =
\begin{cases}
  B^{i,j}_{o} + R^{i,j}_{o,w_{k}} & \text{if } |R^{i,j}_{o,w_{k}}| < |\tilde{R}^{i,j}_{o,w_{1-k}}| \\
  B^{i,j}_{o} + \tilde{R}^{i,j}_{o,w_{1-k}} & \text{if } |R^{i,j}_{o,w_{k}}| \geq |\tilde{R}^{i,j}_{o,w_{1-k}}|,
\end{cases}
\end{equation}

\vspace{-3mm}

\begin{equation}
\label{eq_aow_max}
B^{i,j,k}_{AoW-max} =
\begin{cases}
  B^{i,j}_{o} + \tilde{R}^{i,j}_{o,w_{1-k}} & \text{if } |R^{i,j}_{o,w_{k}}| < |\tilde{R}^{i,j}_{o,w_{1-k}}| \\
  B^{i,j}_{o} + R^{i,j}_{o,w_{k}} & \text{if } |R^{i,j}_{o,w_{k}}| \geq |\tilde{R}^{i,j}_{o,w_{1-k}}|,
\end{cases}
\end{equation}
where $i$ and $j$ denote pixel location and $k$ represents bit information ($0 \leq i < W, 0 \leq j < H, k \in {0,1}$).
AoW-min and AoW-max embed the devised watermark by selectively adding $R^{i}_{o,w_{j}}$ or $\tilde{R}^{i}_{o,w_{1-j}}$ to $B_{o}$.
AoW-min selects the residual with the minimum absolute value, and AoW-max selects with the maximum absolute value to get additional imperceptibility and robustness, respectively (see Algorithm~\ref{algorithm1}).
In detail, when WAN attacks watermarked block with 1 bit embedded $B_{w_1}$, it tries to invert bit to 0 bit with $\tilde{R}_{o,w_1}$ by imitating residual between original block $B_o$ and $B_{w_0}$, which is $R_{o,w_{0}}$.
If attack is succeeded, $\tilde{R}_{o,w_1}$ would be similar to the $R_{o,w_{0}}$.
The motivation of AoW is that $\tilde{R}_{o,w_1}$ can be used for fine-tuning $B_{w_0}$.
AoW simply compares residuals between $\tilde{R}_{o,w_1}$ and $R_{o,w_{0}}$ element-wisely, and AoW-min and AoW-max add the minimum or maximum value to $B_o$, respectively.

With the aid of WAN, which deliberately interferes with MW systems characterized by diverse attributes and intricate processes, our AoW enables significant improvements in the imperceptibility and robustness of the target MW system.
Particularly for rule-based MW, where processes and parameters are empirically optimized, enhancing performance typically requires considerable resource investment.
Nonetheless, our AoW achieves performance gains without necessitating changes to these established rules.

\begin{table*}[t]
\caption{List of attributes and parameters of rule-based MW methods}
\centering
% \scriptsize
\resizebox{\linewidth}{!}{
\begin{tabular}{lcl}
\Xhline{2\arrayrulewidth}
\textbf{MW method} & \textbf{Size of minimum unit} & \textbf{Parameter and value} \\
\hline
\textbf{M1} \cite{kim2018robust} & $1\times64$ (MV) & $N=1$, $M=1$, $\alpha=0.5$, $WM_{len}=20$, $EM_{pos}=15$, $\alpha_{t}=5$, $\beta_{t}=50$ \\
\textbf{M2} \cite{lin2011digital} & $16\times16$ (MV) &  $\alpha=1$, $\lambda=1$, $K_{16\times16}=80$, $EM_{pos}=10$, $t_{x}=0$, \\
\textbf{M3} \cite{su2017improved} & $8\times8$ (MV) &  $T=0.03$, \textit{Size of Matrix} $=3\times3$, $x=y=\{1,2,3\}$\\
\textbf{M4} \cite{makbol2016block} & $8\times8$ (MV) & $T=0.02$, \textit{Size of Matrix} $=4\times4$\\
\multirow{2}{*}{\textbf{M5} \cite{nam2020nsct}}  & \multirow{2}{*}{$-$} &  $N=1$, $M=1$, $\alpha=1$, $\beta=0.7$, $\theta=0.5$, $\epsilon_{1}=0.2$, $\epsilon_{2}=0.8$, $\eta_{1}=0.6$, $\eta_{2}=0.4$,\\
&   & $\mathbb{L}=8$, $\mathbb{S}=2$, $\Delta=2$, $\omega_{1}=\omega_{2}=600$, $\mu=0.1522$ \\
\textbf{M6} \cite{kim2012dtcwt} & $-$ & $N=1$, $M=1$, $errMin=450$, $maxBit=8$, $\mathbb{W}=2$, $TH_1=0.092$, $TH_2=-0.055$\\
\textbf{M7} \cite{parah2016dct} & $16\times16$ (MV) & $E=12$, $T=80$, $V=0.5$, $Length_{lowfreq}=9$\\
% \textbf{M8} \cite{wm_deep2} &  & \\
% \textbf{M9} \cite{zhang2020udh} &  & \\
\Xhline{2\arrayrulewidth}
\multicolumn{3}{l}{$\dagger$ Details of parameters are set based on the notation of each paper.} \\
\multicolumn{3}{l}{$\dagger$ MV is abbreviation of the majority voting that aggregates results of minimum units.} \\
\end{tabular}
}
\label{table_mw_method_para}
\end{table*}

\section{Experiments}
\label{sec_experiments}

\subsection{Settings}
\subsubsection{Configuration Settings of Target MW Methods}
To validate the performance of the proposed WAN and AoW, this study employs handcrafted MW methods performing blind extraction that possess various attributes, such as watermarking domain, embedding algorithm, and key characteristics (see Table~\ref{table_mw_method_para}). 
The attack performance of WAN and the additional enhancement capabilities of AoW on rule-based systems ranging from \textbf{M1} to \textbf{M7} are intensively analyzed in this experimental section \cite{kim2018robust,lin2011digital,su2017improved,makbol2016block,nam2020nsct,kim2012dtcwt,parah2016dct}.
The methods corresponding to \textbf{M1}, \textbf{M2}, \textbf{M3}, \textbf{M4}, and \textbf{M7} embed a watermark in the minimum unit, and aggregate the results extracted from the minimum unit using majority voting on the extraction process.
\textbf{M5} and \textbf{M6}, lacking a fixed minimum unit size, adaptively divide the input based on the capacity to determine variable unit sizes.
Details of parameter setting of each MW method are listed in Table~\ref{table_mw_method_para}, and key summaries of each method are as follows.

\vspace{-2mm}
\begin{itemize}
\item \textbf{M1} \cite{kim2018robust} is designed to robust against desynchronization attacks.
This method embeds the watermarks in the 1D DCT coefficients of each minimum unit ($1\times64$) based on SS embedding.

\item \textbf{M2} \cite{lin2011digital} embeds multiple watermarks, which are designed to minimize interference between each watermark, into minimum unit of size $16\times16$.
The method applies DCT to each minimum unit, and then inserts a watermark into the coefficients using ISS embedding.

\item \textbf{M3} \cite{su2017improved} first obtain non-overlapping $3\times3$ pixels from minimum unit ($8\times8$), and then QRD is applied to obtained pixels.
This method embeds a watermark by adjusting the relation between the coefficients located in the first column of the second and third rows of orthogonal matrix Q.

\item \textbf{M4} \cite{makbol2016block} selects significant minimum units ($8\times8$) based on entropy map as HVS characteristics.
After first-level DWT decomposition on the selected units, a watermark is inserted by applying subtle deformation to the U matrix of the SVD according to a predefined condition.

\item \textbf{M5} \cite{nam2020nsct} uses two-level NSCT decomposition and embeds watermark in the low coefficients of NSCT subbands using QT embedding for ensuring robustness against pixel-level shift.
It improves the imperceptibility by adjusting the embedding strength based on computed perceptual masking value.

\item \textbf{M6} \cite{kim2012dtcwt} adopts DT-CWT, which possesses characteristics of approximate shift invariance and directional selectivity, to secure robustness against geometric attacks.
This method aims to enhance imperceptibility by selectively modifying specific coefficients within the DTCWT subbands during the embedding process.

\item \textbf{M7} \cite{parah2016dct} introduces $16\times16$ unit block-based DCT coefficient modification for robust watermarking.
It calculates the difference between two DCT coefficients of adjacent blocks at the same position and adjusts this difference to fall within a predefined range by modifying one of the coefficients.
\end{itemize}

\begin{table*}[t]
\caption{Details of grey-scale image datasets}
\centering
% \scriptsize
\resizebox{\linewidth}{!}{
\begin{tabular}{lccccccccc}
\Xhline{2\arrayrulewidth}
\multirow{3}{*}{\textbf{MW method}} & \multicolumn{9}{c}{\textbf{Number of block images}}\\
\cmidrule(r){2-10}
& \multicolumn{3}{c}{\textbf{1 bit (64$\times$64)}} & \multicolumn{3}{c}{\textbf{4 bits (128$\times$128)}} & \multicolumn{3}{c}{\textbf{16 bits (256$\times$256)}} \\
\cmidrule(r){2-4} \cmidrule(r){5-7} \cmidrule(r){8-10}
 &  $B_{o}$ &  $B_{w_{0}}$ &  $B_{w_{1}}$ &  $B_{o}$ &  $B_{w_{0}}$ &  $B_{w_{1}}$ &  $B_{o}$ &  $B_{w_{0}}$ &  $B_{w_{1}}$ \\
\hline
\textbf{M1} \cite{kim2018robust} & 20K & 20K & 20K & 5K & 5K & 5K & 5K & 5K & 5K  \\
\textbf{M2} \cite{lin2011digital} & 20K & 20K & 20K & 5K & 5K & 5K & 5K & 5K & 5K  \\
\textbf{M3} \cite{su2017improved} & 20K & 20K & 20K & 5K & 5K & 5K & 5K & 5K & 5K  \\
\textbf{M4} \cite{makbol2016block} & 20K & 20K & 20K & 5K & 5K & 5K & 5K & 5K & 5K  \\
\textbf{M5} \cite{nam2020nsct} & 20K & 20K & 20K & 5K & 5K & 5K & 5K & 5K & 5K  \\
\textbf{M6} \cite{kim2012dtcwt} & 20K & 20K & 20K & 5K & 5K & 5K & 5K & 5K & 5K  \\
\textbf{M7} \cite{parah2016dct} & 20K & 20K & 20K & 5K & 5K & 5K & 5K & 5K & 5K  \\
\hline
Total & 140K & 140K & 140K & 35K & 35K & 35K & 35K & 35K & 35K  \\
\Xhline{2\arrayrulewidth}
\end{tabular}
}
\label{table_dataset}
\end{table*}

\subsubsection{Datasets}
\label{sec_dataset}

In this study, we utilize BOSSbase \cite{boss} and BOWS \cite{bows} datasets, which are widely used in the field of \textit{watermarking and steganography}, to generate 20,000 original grey-scale images with a size of $512\times512$.
For base experiment, we resize them to $64\times64$ (i.e., $W=H=64$) using the default settings in MATLAB R2018a, and the resized images are divided into three sets for training, validation, and test (with a $14:1:5$ ratio). 
The handcrafted MW methods (\textbf{M1} $-$ \textbf{M7}) are employed to generate watermarked images, and the images are generated by embedding watermark bits (0 or 1) into the original images given for each method.
In experiments, training of the WAN for the target MW is conducted using a training dataset (resolution of $64\times64$) with a 1 bit capacity, and the WAN-based attacks and watermark bit extraction proceeds for each $64\times64$ patch.
For further quantitative and qualitative evaluation on watermark capacity, we additionally generate test images sized $128\times128$ and $256\times256$ for the test set.
Watermarked images with resolutions of $128\times128$ and $256\times256$ have a watermark capacity of 4 bits and 16 bits, respectively.
Experiments on a test set with multi-bit insertion capacity are conducted using a pre-trained WAN model with a stride of $64$.
The detailed description of dataset is listed in Table~\ref{table_dataset}.
Details on block images (e.g., $\tilde{B}_{w_{0}}$ or $\tilde{B}_{w_{0}}$) based on WAN's output are excluded from the table.

\subsubsection{Implementation Details and Training Settings}
\label{sec_training_settings}
%We construct our model based on the CNN components shown in Section~\ref{sec_architecture}. 
The number of RDB, Conv layer per RDB, feature-maps, and the growth rate are set to 12, 6, 32, and 16, respectively.
We build our network using PyTorch and run the experiments on NVIDIA GeForce GTX 1080 Ti.
The size of mini-batch $b_s$ is set to 32, and each mini-batch is configured for paired mini-batch training \cite{paired}.
We use the Adam optimizer with a learning rate of $10^{-4}$ and momentum coefficients $\beta_{1}=0.9$, $\beta_{2}=0.999$.
The proposed WAN is trained with the hyperparameters $\lambda_{c}=0.4$ and $\lambda_{wa}=0.3$ during 30 epochs, and the best model is selected as the one that maximizes BER on the validation set for each MW method.

\subsubsection{Evaluation Metrics}
\label{sec_metric}

The proposed WAN is designed to induce abnormal extraction in the target MW while minimizing visual degradation.
To evaluate the enhanced perceptual quality compared to traditional attacks, we use IQA metrics, PSNR (dB) and SSIM \cite{ssim}.
Additionally, to assess whether the application of WAN has successfully degraded the watermark extraction performance as intended (indicating a successful attack from the watermarking attack perspective), BER is used as an evaluation metric.
The BER between n-bit-long original message ($\mathbf{m}=\{0,1\}^n$) and extracted message $\mathbf{\hat{m}}$ (i.e., the message extracted by the watermark extractor after applying WAN to the watermarked content) is defined as follows: $\text{BER}(\textbf{m}, \mathbf{\hat{m}}) = \frac{\text{number of } (\mathbf{b_i} \text{ in } \mathbf{m} \neq \mathbf{\hat{b_i}} \text{ in } \mathbf{\hat{m}})}{n} \text{ for } 0 \leq i \leq n - 1$. 
In the case where $n$ is set to 1, this corresponds to a watermark capacity of 1 bit; therefore, $\mathbf{\hat{m}}$ and $\mathbf{\hat{b}}$ are equal.
Additionally, since AoW aims to enhance the imperceptibility and robustness of the target handcrafted MW, it is quantitatively assessed using the same metrics.

\begin{table*}[t]
\caption{Quantitative evaluation results of the proposed WAN on the test set with 1 bit, 4 bits, and 16 bits watermark capacities}
\resizebox{\linewidth}{!}{
\begin{tabular}{ccccccccccccccccccc}
\Xhline{2\arrayrulewidth}
& \multicolumn{6}{c}{\textbf{1 bit of watermark capacity}} & \multicolumn{6}{c}{\textbf{4 bits of watermark capacity}}  & \multicolumn{6}{c}{\textbf{16 bits of watermark capacity}}\\ 
\cmidrule(r){2-7} \cmidrule(r){8-13} \cmidrule(r){14-19}
\multicolumn{1}{c}{\textbf{MW}} & \multicolumn{3}{c}{\textbf{Non-attack}} & \multicolumn{3}{c}{\textbf{WAN}} & \multicolumn{3}{c}{\textbf{Non-attack}} & \multicolumn{3}{c}{\textbf{WAN}} & \multicolumn{3}{c}{\textbf{Non-attack}} & \multicolumn{3}{c}{\textbf{WAN}} \\
\cmidrule(r){2-4} \cmidrule(r){5-7} \cmidrule(r){8-10} \cmidrule(r){11-13} \cmidrule(r){14-16} \cmidrule(r){17-19}
& PSNR & SSIM & BER & PSNR & SSIM & BER & PSNR & SSIM & BER & PSNR & SSIM & BER & PSNR & SSIM & BER & PSNR & SSIM & BER \\
\hline
\textbf{M1} & 35.55 & 0.938 & 0.026 & 34.04 & 0.956 & 0.893 & 35.53 & 0.938 & 0.049 & 34.79 & 0.963 & 0.905 & 37.73 & 0.957 & 0.043 & 35.98 & 0.971 & 0.882 \\
\textbf{M2} & 41.86 & 0.988 & 0 & 37.47 & 0.979 & 0.996 & 42.62 & 0.987 & 0  & 37.44 & 0.980 & 0.993 & 43.11 & 0.985 & 0 & 38.30 & 0.982 & 0.990\\
\textbf{M3} & 36.59 & 0.974 & 0 & 33.05 & 0.96 & 1.000 & 37.55 & 0.973 & 0 & 33.64 & 0.962 & 0.998 & 38.33 & 0.973 & 0 & 35.31 & 0.976 & 0.994\\
\textbf{M4} & 38.98 & 0.986 & 0.002 & 37.70 & 0.985 & 0.988 & 39.71 & 0.985 & 0.002 & 38.09 & 0.985 & 0.990 & 39.88 & 0.979 & 0.003 & 38.44 & 0.980 & 0.991\\
\textbf{M5} & 39.21 & 0.987 & 0.013 & 36.54 & 0.980 & 0.947 & 40.64 & 0.987 & 0.041 & 37.22 & 0.982 & 0.885 & 41.63 & 0.989 & 0.032 & 38.48 & 0.985 & 0.851\\
\textbf{M6} & 36.61 & 0.972 & 0.021 & 35.52 & 0.970 & 0.912 & 37.52 & 0.973 & 0.044 & 35.85 & 0.971 & 0.874 & 37.89 & 0.975 & 0.052 & 36.12 & 0.976 & 0.842\\
\textbf{M7} & 36.43 & 0.945 & 0 & 35.62 & 0.938 & 0.890 & 36.41 & 0.942 & 0 & 35.93 & 0.941 & 0.881 & 36.48 & 0.944 & 0 & 35.77 & 0.942 & 0.872\\
\hline
Average & 37.89 & 0.970 & 0.009 & 35.71 & 0.967 & 0.946 & 38.57 & 0.969 & 0.019 & 36.13 & 0.969 & 0.932 & 39.29 & 0.972 & 0.019 & 36.91 & 0.973 & 0.917\\
% Average & 38.44 & 0.974 & 0.008 & 35.76 & 0.972 & 0.965 & 39.21 & 0.974 & 0.018 & 36.24 & 0.974 & 0.954 & 40.14 & 0.977 & 0.015 & 37.30 & 0.978 & 0.942 \\

% 37.89	0.97	0.008857143	35.70571429	0.966857143	0.946571429	38.56857143	0.969285714	0.019428571	36.13714286	0.969142857	0.932285714	39.29285714	0.971714286	0.018571429	36.91428571	0.973142857	0.917428571

\Xhline{2\arrayrulewidth}
\end{tabular}}
\label{table_quantitative_evaluation}
\vspace{-5mm}
\end{table*}

\subsection{Performance Evaluation of WAN}

\subsubsection{Quantitative Evaluation of WAN}

First, a quantitative evaluation of the WAN is conducted in terms of watermark extraction interference and the visual quality of attacked images.
The left part of Table~\ref{table_quantitative_evaluation} shows the performance results of our work on the test set with 1 bit capacity generated through each MW method (\textbf{M1} $-$ \textbf{M7}), which are composed of various attributes.
In non-attack situations, each method has a low BER value of 0.026 or less, while the average BER value increases dramatically to 0.946 after WAN is applied.
In particular, for MW methods in \cite{lin2011digital,su2017improved,makbol2016block}, the BER value of methods rise to 0.988 or more, which means that the WAN has learned a fine signal generated during the watermark embedding and successfully performs bit inversion.
In general, making the extraction performance at a random guessing level is considered a very fatal attack \cite{wm_base6}, and it is validated that the proposed $\mathcal{L}_{wa}$ successfully leads to abnormal extraction of watermark bits.
Additionally, the performance of WAN attacks across various techniques exhibits minor discrepancies, which are assumed to originate from the varied attributes used in watermarking design and the differing embedding strengths with parameter setting.
This observation implies that the vulnerabilities and the complexities associated with attacks differ among techniques, suggesting that WAN can function effectively as a benchmarking tool in this context.

In addition, minimizing the visual degradation caused by watermarking attacks is an important issue in our work.
To do this, we introduce $\mathcal{L}_{c}$, and the gain of visual quality obtained from the loss can be analyzed through PSNR and SSIM values with original content in Table~\ref{table_quantitative_evaluation}.
In case of 1 bit, the average PSNR and SSIM values in non-attack situation are 37.89 dB and 0.970, respectively.
After the WAN attacks images, average PSNR decreases by 2.18 dB, and average SSIM decreases by 0.003.
We would like to note that existing benchmark tools \cite{stirmark1,checkmark} have to degrade images up to approximately 20 dB for noise addition to get random guessed results, as shown in Figure~\ref{figure_visual_degradation_stirmark}.
Meanwhile, our model is capable of inducing the drastic reversal of the watermark bit with acceptable small loss of perceptual quality.
Additionally, the visualized residual images in Figure~\ref{figure_attack_example_intro} further confirm that the proposed WAN conducts attacks with less distortion compared to other watermarking attacks.
As indicated in the table, following the application of the attack, the average PSNR and SSIM are recorded as 35.71 dB and 0.967, respectively, demonstrating from a quantitative evaluation perspective an improvement in perceptual quality compared to previous attacks.

We further test for 4 and 16 bits of watermark capacity scenarios with the trained WAN model with stride $64$ (see Section~\ref{sec_dataset})
As listed in the middle side of Table~\ref{table_quantitative_evaluation}, for 4 bits capacity, the average PSNR, SSIM, and BER values for the attacked images over the WAN are 36.13 dB, 0.969, and 0.932, respectively.
For the results of the 16 bits capacity, the average BER value is 0.917 while achieving the improved visual quality.
% Compared with the results for 1 bit capacity, we can confirm that the WAN's overall performance is maintained even when the watermark capacity is increased.
Compared with the results for 1 bit capacity, the WAN's performance slightly declines when the watermark capacity is increased; however, this still represents high performance from an attack perspective, as it maintains imperceptibility while inducing abnormal extraction.
This slight decline is presumed to stem from differences in content representation due to resolution discrepancies between the training and testing phases.
Overall, the results of quantitative evaluation show that the proposed WAN is suitable for testing MW methods as a benchmark tool in terms of interference of watermark extraction, maintenance of visual quality, and scalability according to watermark capacity.

\begin{figure*}[t]
\centering{\includegraphics[width=1.0\linewidth]{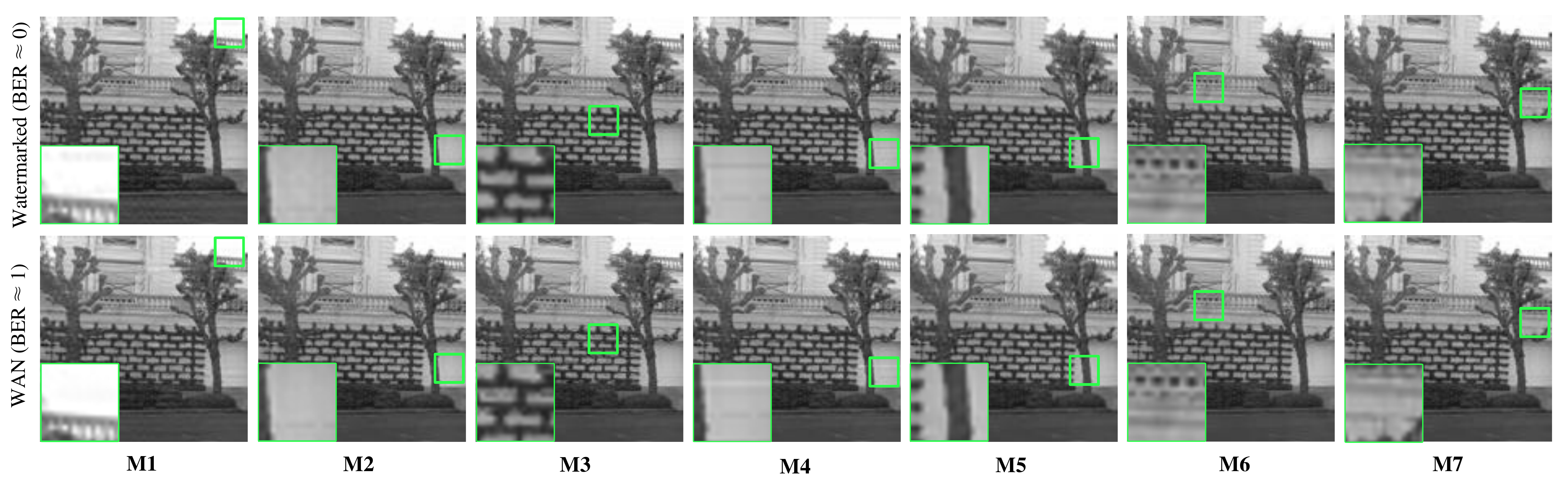}}
\caption{Examples of attacked images generated from the WAN applied to the watermarked image with 4 bits of capacity.}
\label{figure_qualitative_result}
\end{figure*}

\subsubsection{Qualitative Evaluation of WAN}
\label{sec_qualitative_evaluation}

In this section, we perform qualitative evaluation in terms of perceptual quality.
Figure~\ref{figure_qualitative_result} shows the examples of the watermarked image with 4 bits of capacity and the attacked image of the proposed WAN.
% As shown in the top row of Figure~\ref{figure_qualitative_result}, the types of low-level distortion caused by watermark embedding vary by MW method while having similar high-level features (i.e., inherent content).
As shown in the figure, the types of low-level distortion caused by watermark embedding vary by MW method while having similar high-level features (i.e., inherent content).
For MW systems \cite{kim2018robust,lin2011digital,su2017improved,makbol2016block,nam2020nsct,kim2012dtcwt,parah2016dct}, the WAN hardly causes visual degradation in the process of inverting watermark bits (see magnified sub-figures in Figure~\ref{figure_qualitative_result}).
The proposed WAN with $\mathcal{L}_{wa}$ and $\mathcal{L}_{c}$ can hinder watermark extraction by learning these fine feature and induces the attacked image to visually follow the original content.
% Our work can produce natural attacked results that very similar to images in non-attack situations (see bottom row of Figure~\ref{figure_qualitative_result}).

\begin{figure*}[t]
\centering{\includegraphics[width=0.5\linewidth]{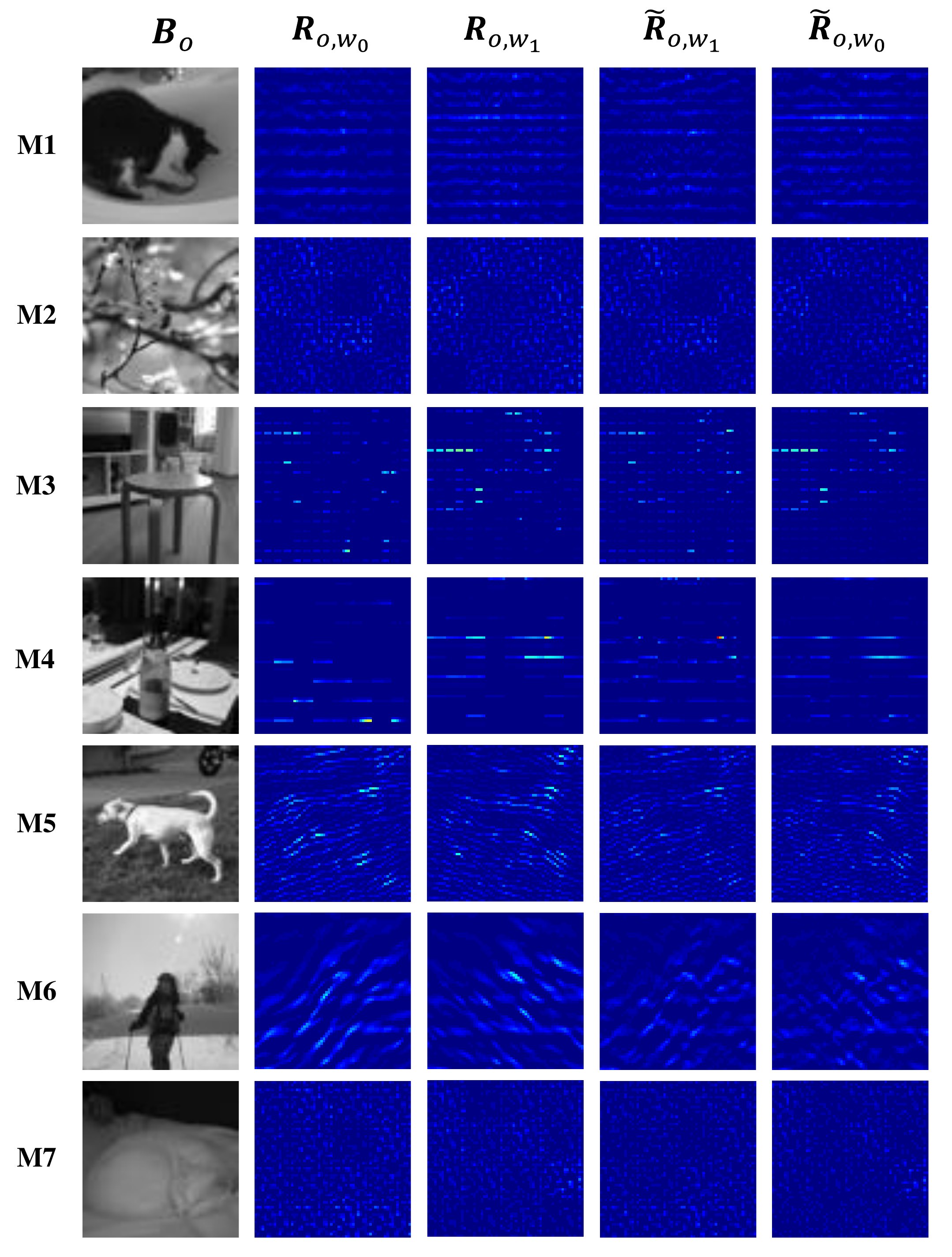}}
\caption{Visualization of residual images. $\tilde{R}_{o,w_{1}}$ and $\tilde{R}_{o,w_{0}}$ are reconstructed from $R_{o,w_{0}}$ and $R_{o,w_{1}}$, respectively.}
\label{figure_residual_image_analysis}
\end{figure*}

% \begin{table*}[t]
% \caption{Ablation study results of the proposed WAN on loss function}
% \scriptsize
% \centering
% \begin{tabular}{cccccccccccccccc}
% \Xhline{2\arrayrulewidth}
% \multirow{3}{*}{\shortstack{\textbf{Loss}}} & \multicolumn{3}{c}{\textbf{M1}} & \multicolumn{3}{c}{\textbf{M2}} & \multicolumn{3}{c}{\textbf{M3}} & \multicolumn{3}{c}{\textbf{M4}} & \multicolumn{3}{c}{\textbf{M5}}\\
% \cmidrule(r){2-4}\cmidrule(r){5-7}\cmidrule(r){8-10}\cmidrule(r){11-13}\cmidrule(r){14-16}
% & PSNR & SSIM & BER & PSNR & SSIM & BER & PSNR & SSIM & BER & PSNR & SSIM & BER & PSNR & SSIM & BER\\
% \hline
% $\mathcal{L}_{c}$ & 38.49 & 0.983 & 0.526 & 43.94 & 0.995 & 0.497 & 41.79 & 0.992 & 0.513 & 42.5 & 0.994 & 0.525 & 41.93 & 0.991 & 0.464 \\
% $\mathcal{L}_{wa}$ & 30.38 & 0.905 & 0.914 & 35.55 & 0.964 & 0.996 & 31.48 & 0.947 & 1 & 34.88 & 0.972 & 0.997 & 34.17 & 0.969 & 0.972 \\
% $\mathcal{L}$ & 34.04 & 0.956 & 0.893 & 37.47 & 0.979 & 0.996 & 33.05 & 0.96 & 1 & 37.7 & 0.985 & 0.988 & 36.54 & 0.98 & 0.947 \\
% \Xhline{2\arrayrulewidth}
% \end{tabular}
% \vspace{-4mm}
% \label{table_ablation_study}
% \end{table*}

%Next, we acquire four types of residual images $(\{R_{o,w_{0}},R_{o,w_{1}},\tilde{R}_{o,w_{1}},\tilde{R}_{o,w_{0}}\}$) and analyze WAN-based attacks by comparing the differences between residual images.
Next, to more thoroughly analyze WAN-based attacks, we acquire four types of residual images $(\{R_{o,w_{0}},R_{o,w_{1}},\tilde{R}_{o,w_{1}},\tilde{R}_{o,w_{0}}\}$) by comparing the differences between them.
As described in Section~\ref{sec_watermarking_attack_loss}, the WAN is guided to reduce the difference between the paired residual images during the training phase, thereby applying an adversarial signal to $B_w$ that causes abnormal extraction.
High similarity between $R_{o,w_{0}}$ and $\tilde{R}_{o,w_{1}}$ located in the 2$nd$ and 4$th$ columns in Figure~\ref{figure_residual_image_analysis} is observed, indicating that the WAN successfully attacks the watermarked image containing watermark bit 1 (i.e., 1$\rightarrow$0).
Similarly, the characteristics of residual images in the 3$rd$ and 5$th$ columns are co-related.
% The small difference between the 3$rd$ and 5$th$ columns means that the WAN reduce the difference between the attacked image and the watermarked image with watermark bit 1.
Furthermore, since MW methods with various attributes are used in this study, the distribution and characteristics of each residual data vary depending on the methods.
% From this, we can confirm that WAN, when trained to learn the properties (e.g., the watermarking domain and embedding method) of each MW, analyzes weak points of MW system adaptively.
In summary, we can confirm that WAN can adaptively learn the individual characteristics of the MW method, such as watermarking domain and embedding algorithm.

Figure~\ref{figure_visual_degradation_stirmark} compares results of the our model and StirMark \cite{stirmark1,stirmark2} consisting signal processing operations and geometric distortions.
For fairness in comparison, attacked images generated through attack parameters of StirMark that cause random guessing of bit extraction (e.g., BER $\approx$ 0.5) are compared.
As mentioned above, the StirMark is not an approach of attacking by modeling the vulnerability of the MW method or considering inherent content, so it is accompanied by unwanted visual degradation in the attack process (see magnified sub-figures in Figure~\ref{figure_visual_degradation_stirmark}).
In contrast, our WAN can adversely affect the extraction of the inserted bit while maintaining the inherent properties of the original content.
From the results of qualitative evaluation, it is confirmed that the network architecture specialized for image restoration and the proposed loss function are effective in generating natural attacked images.

\begin{table*}[t]
\caption{Ablation study results of the proposed WAN on loss function}
% \scriptsize
\resizebox{\linewidth}{!}{
\centering
\begin{tabular}{lccccccccc}
\Xhline{2\arrayrulewidth}
\multirow{2}{*}{\textbf{MW}} & \multicolumn{3}{c}{\textbf{$\mathcal{L}_{c}$}} & \multicolumn{3}{c}{\textbf{$\mathcal{L}_{wa}$}} & \multicolumn{3}{c}{\textbf{$\mathcal{L}$}} \\
\cmidrule(r){2-4}\cmidrule(r){5-7}\cmidrule(r){8-10}
& PSNR & SSIM & BER & PSNR & SSIM & BER & PSNR & SSIM & BER \\
\hline
\textbf{M1} \cite{kim2018robust} & 38.49 & 0.983 & 0.526 & 30.38 & 0.905 & 0.914 & 34.04 & 0.956 & 0.893 \\
\textbf{M2} \cite{lin2011digital} & 43.94 & 0.995 & 0.497 & 35.55 & 0.964 & 0.996 & 37.47 & 0.979 & 0.996 \\
\textbf{M3} \cite{su2017improved} & 41.79 & 0.992 & 0.513 & 31.48 & 0.947 & 1 & 33.05 & 0.96 & 1 \\
\textbf{M4} \cite{makbol2016block} & 42.5 & 0.994 & 0.525 & 34.88 & 0.972 & 0.997 & 37.7 & 0.985 & 0.988 \\
\textbf{M5} \cite{nam2020nsct} & 41.93 & 0.991 & 0.464 & 34.17 & 0.969 & 0.972 & 36.54 & 0.98 & 0.947 \\
\textbf{M6} \cite{kim2012dtcwt} & 40.14 & 0.987 & 0.496 & 31.36 & 0.943 & 0.936 & 35.52 & 0.970 & 0.912 \\ 
\textbf{M7} \cite{parah2016dct} & 40.22 & 0.988 & 0.508 & 31.28 & 0.915 & 0.910 & 35.62 & 0.938 & 0.890 \\ 
\hline
\textbf{Average} & 41.29 & 0.990 & 0.504 & 32.73 & 0.945 & 0.961 & 35.71 & 0.967 & 0.946 \\ 
\Xhline{2\arrayrulewidth}
\end{tabular}
}
\label{table_ablation_study}
\end{table*}

\subsection{Comprehensive Analysis of WAN}
In this section, we present the results of extended experiments to conduct a comprehensive analysis of WAN.

\subsubsection{Ablation Study on Loss Function}
\label{sec_ablation_study}

To investigate how loss function $\mathcal{L}$ contributes to the overall performance of our WAN, we conduct an ablation study on the test set with 1 bit of watermark capacity.
We perform experiments on the following losses: content loss $\mathcal{L}_{c}$, watermarking attack loss $\mathcal{L}_{wa}$, and combined $\mathcal{L}$.
The results are summarized in Table~\ref{table_ablation_study}.
% For $\mathcal{L}_{c}$ only, each MW shows outstanding performance in terms of fidelity, but the effectiveness of WAN decreases to the level of random guessing.
For $\mathcal{L}_{c}$ only, each MW demonstrates outstanding performance in terms of fidelity, achieving average values of PSNR 41.29 dB and SSIM 0.990; however, the effectiveness of WAN decreases to the level of random guessing, with an average BER of 0.504.
When training $\mathcal{L}_{wa}$ only, all MW methods show a high BER value (of 0.910 or higher), but the visual quality is greatly reduced during the attack process.
In particular, $\mathcal{L}_{wa}$-based WAN shows a tendency to rapidly improve attack performance with a BER value of 0.85 or more before 10 epochs.
The results presented in Table~\ref{table_ablation_study} demonstrate that the watermarking attack loss $\mathcal{L}_{wa}$ and content loss $\mathcal{L}_{c}$ contribute to achieving sophisticated watermarking attacks (i.e., watermark bit inversion) and preserving perceptual quality, respectively.
As described in Section~\ref{sec_loss}, the proposed objective function $\mathcal{L}$ is designed considering the advantages of $\mathcal{L}_{c}$ and $\mathcal{L}_{wa}$ jointly.
From the results in this section, we can conclude that $\mathcal{L}$ helps the model to leverage the advantages of $\mathcal{L}_{c}$ and $\mathcal{L}_{wa}$ simultaneously.

\begin{figure*}[t]%
\centering%
\subfigure[]{%
  \includegraphics[width=0.4\linewidth]{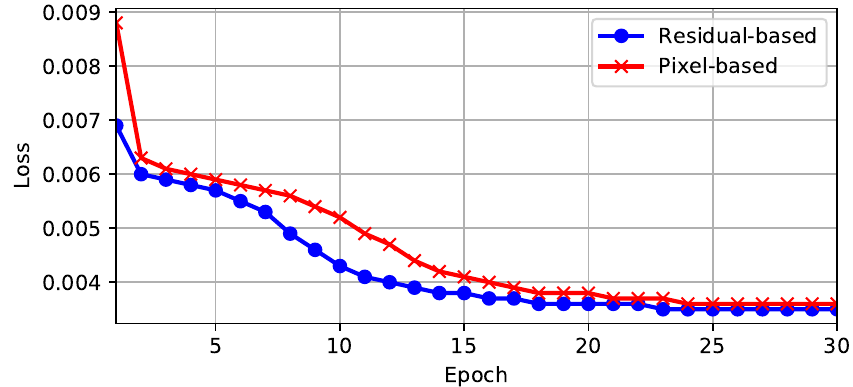}
  \label{fig_tendency:a}
}
\subfigure[]{%
  \includegraphics[width=0.4\linewidth]{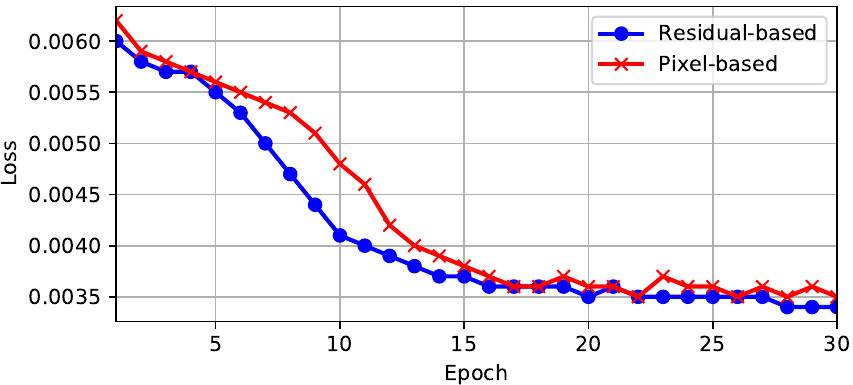}
  \label{fig_tendency:b}
}
\subfigure[]{%
  \includegraphics[width=0.4\linewidth]{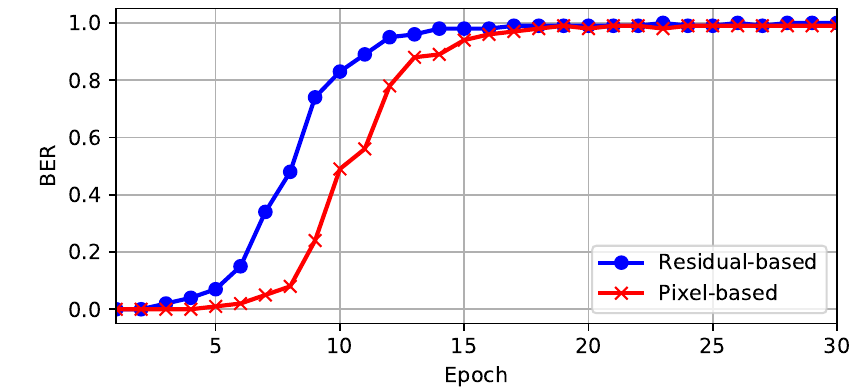}
  \label{fig_tendency:c}
}
\subfigure[]{%
  \includegraphics[width=0.4\linewidth]{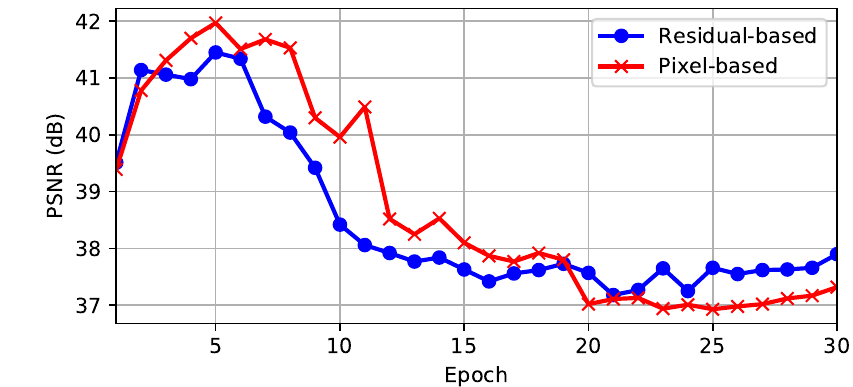}
  \label{fig_tendency:d}
}

\caption{Training loss (a), validation loss (b), validation BER (c), and validation PSNR (d) tendencies for each network over 30 epochs. The experiment was conducted based on \textbf{M2}.}
\label{fig_tendency}%
\end{figure*}

\subsubsection{Impact of Input Format on Loss Function}
\label{residual_data}

In this section, we present the training tendencies and performance analysis based on the input format of watermarking attack loss $\mathcal{L}_{wa}$ of the proposed WAN.
As defined in Equation~(\ref{eq1}), the proposed WAN induces abnormal watermark extraction (i.e., watermark bit inversion) based on residual data.
To evaluate the effectiveness of the residual-based approach, we perform a performance analysis and compare it with pixel-based watermarking loss $\mathcal{L}^{p}_{wa}$, which leverages the difference between images.
To this end, $\mathcal{L}^{p}_{wa}$ is defined as follows : $\mathcal{L}^{p}_{wa} = \frac{1}{N}\sum_{i=1}^{N}|B^{i}_{w_0}-\tilde{B}^{i}_{w_1}| + \frac{1}{N}\sum_{i=1}^{N}|B^{i}_{w_1}-\tilde{B}^{i}_{w_0}|$.
When the watermarking attack loss is changed from residual differences (e.g., $|R_{o,w_{0}}-\tilde{R}_{o,w_{1}}|$) into pixel differences (e.g., $|B_{w_0}-\tilde{B}_{w_1}|$), we found that the speed of convergence gets slower and that the effectiveness of watermarking attack degrades (see Figure~\ref{fig_tendency}).
This is because the scale and variance of outputs with pixel differences become bigger than the ones with residual differences, which makes WAN training more difficult.
Additionally, it is observed that the residual-based approach shows stable validation PSNR and validation BER tendencies.
Based on the results analyzed in this section, it can be observed that the objective function based on residual difference (i.e., $\mathcal{L}_{wa}$) proposed in our work is more suitable for model training and performance enhancement than the approach based on pixel difference.

\begin{table}[t]
\caption{Comparison of performance with the baseline}
\scriptsize
\centering
\begin{tabular}{lcccccc}
\Xhline{2\arrayrulewidth}
 & \multicolumn{3}{c}{WAN w/ \textbf{$\mathcal{L}_{c}$}} & \multicolumn{3}{c}{FCNNDA}  \\
\cmidrule(r){2-4}\cmidrule(r){5-7}
& PSNR & SSIM & BER & PSNR & SSIM & BER\\
\hline
\textbf{Average} & 41.29 & 0.990 & 0.504 & 37.68 & 0.978 & 0.485 \\ 
\Xhline{2\arrayrulewidth}
\end{tabular}
\label{table_baseline}
\end{table}

\subsubsection{Comparison with baseline}
As introduced in Section~\ref{sec_watermarking_attack_motivation} and Section~\ref{sec_proposed_method}, our WAN aims to explore the vulnerabilities of the target MW system, inducing abnormal extraction by inverting watermark bits (i.e., BER $\approx$ 1).
As verified in Section~\ref{sec_ablation_study}, when trained solely with content loss, the WAN can reconstruct attacked images to closely match the distribution of the original content, resulting in a random guessing level of attack effectiveness (i.e., BER $\approx$ 0.5).
Recently, CNN-based attacks have been proposed, particularly FCNNDA~\cite{hatoum2021using}, which involves removing the watermark signal and reconstructing the content to closely resemble the original, thereby disrupting watermark extraction.
This section conducts a comparative analysis of the performance between the WAN, trained using content loss $\mathcal{L}_{c}$, and FCNNDA, which serves as the baseline model.
As listed in Table~\ref{table_baseline}, it has been observed that both the WAN trained with content loss and the baseline model guide the attacked images to follow the distribution of natural images, resulting in abnormal extractions that approximate random guessing.
On the other hand, in terms of imperceptibility, the WAN exhibits higher performance than the baseline, surpassing it by 3.61 dB in PSNR and by 0.012 in SSIM.
This indicates that the proposed WAN effectively carries out attacks while reducing degradation caused by the attacks.

\begin{table}
\caption{Quantitative evaluation results of the proposed WAN targeting the performance of CNN-based MW system with a capacity of 1 bit}
\scriptsize
\centering
\begin{tabular}{lcccccc}
\Xhline{2\arrayrulewidth}
\multirow{3}{*}{\shortstack{\textbf{MW}}} & \multicolumn{3}{c}{\textbf{Non-attack}} & \multicolumn{3}{c}{\textbf{WAN}} \\
\cmidrule(r){2-4} \cmidrule(r){5-7} 
& PSNR & SSIM & BER & PSNR & SSIM & BER \\
\hline
\textbf{M8}  & 41.34 & 0.984 & 0 & 38.26 & 0.975 & 0.920 \\
\Xhline{2\arrayrulewidth}
\end{tabular}
\label{table_results_CNN_MW}
\end{table}

\subsubsection{Performance of WAN on CNN-based MW}

The authors of \cite{wm_deep2} proposed WMNet (\textbf{M8}), which adaptively secures robustness against various attacks using an attack simulator and demonstrates high imperceptibility and robustness.
In this section, we validate the attack performance of WAN against the CNN-based watermarking approach designated as \textbf{M8}.
As mentioned in Table~\ref{table_dataset} and Section~\ref{sec_dataset}, 20,000 original gray-scale blocks of $64\times64$ resolution were divided with a 14:1:5 ratio for the training, validation, and testing of \textbf{M8}.
In detail, the parameters \(\alpha\) and \(\eta\) were set to 0.01 and 0.0004, respectively. Due to the input having only one channel, the network structure was accordingly modified. Other training methodologies followed those specified in the paper~\cite{wm_deep2}, and the attack simulator included noising, Gaussian blur, rotation, crop, rescaling, median blur, and JPEG compression \cite{kwon2025safire}.
In the configuration depicted in the lower part of Figure~\ref{figure_multibit_watermarking_CNN}, \textbf{M8} functions by adding a watermark pattern $P$ to the feature maps generated as the original image passes through the encoder. For the purposes of this experiment, to ensure a watermarking capacity of one bit, the pattern $P$ was consistently set as either a black background image (for watermark bit 0) or a white background image (for watermark bit 1).

Upon completing the training phase for \textbf{M8}, original blocks $B_{o}$ are processed through the trained \textbf{M8} model to produce watermarked block images (i.e., $B_{w_{0}}$ and $B_{w_{1}}$).
These images are subsequently utilized for the training, validation, and testing phases of the WAN.
As listed in Table~\ref{table_results_CNN_MW}, in non-attack scenarios, \textbf{M8} exhibits a low BER value of 0; however, the BER value increases dramatically to 0.920 when WAN is applied.
After the WAN attacks images, the average PSNR decreases by 3.08 dB, and the average SSIM decreases by 0.009.
It is important to note that, WAN effectively conducts attacks on the CNN-based MW system while maximally preserving the perceptual quality.

\begin{table*}
\caption{Quantitative evaluation results of the proposed WAN on color images with 1 bit of capacity}
\scriptsize
\centering
\begin{tabular}{lcccccc}
\Xhline{2\arrayrulewidth}
\multirow{3}{*}{\shortstack{\textbf{MW method}}} & \multicolumn{3}{c}{\textbf{Non-attack}} & \multicolumn{3}{c}{\textbf{WAN}} \\
\cmidrule(r){2-4} \cmidrule(r){5-7} 
& PSNR & SSIM & BER & PSNR & SSIM & BER \\
\hline
\textbf{M1} \cite{kim2018robust} & 40.63 & 0.968 & 0.048 & 37.14 & 0.956 & 0.855  \\
\textbf{M2} \cite{lin2011digital} & 43.14 & 0.979 & 0 & 39.51 & 0.971 & 0.942  \\
\textbf{M3} \cite{su2017improved} & 40.89 & 0.972 & 0 & 38.19 & 0.963 &  0.987  \\
\textbf{M4} \cite{makbol2016block} & 41.03 & 0.969 & 0 & 38.41 & 0.965 & 0.977  \\
\textbf{M5} \cite{nam2020nsct} & 43.07 & 0.988 & 0.037 & 41.97 & 0.981 & 0.902  \\
\textbf{M6} \cite{kim2012dtcwt} & 40.94 & 0.970 & 0.040 & 38.44 & 0.965 & 0.886\\
\textbf{M7} \cite{parah2016dct} & 40.88 & 0.971 & 0 & 37.85 & 0.960 & 0.866 \\
\hline
Average & 41.51 & 0.973 & 0.018 & 38.79 & 0.966 & 0.916\\
\Xhline{2\arrayrulewidth}
\end{tabular}
\label{table_results_color_image}
\end{table*}

\subsubsection{Results on Color Dataset}

In this subsection, we analyze the performance of the proposed WAN on a color dataset.
To do this, color image dataset \cite{park2018double} consisting of single-compressed images based on RAISE \cite{dang2015raise} and Dresden \cite{gloe2010dresden} are exploited \footnote{https://github.com/plok5308/DJPEG-torch}.
Based on random sampling, we select 20,000 original color images with a size of $256\times256$.
We resize them to $64\times64$ (i.e., ${3\times 64\times 64}$) using the default settings in MATLAB R2018a, and the resized images are divided into three sets for training, validation, and test (with a $14:1:5$ ratio).
The number of the generated color images is the same as that of the gray-scale image with 1 bit of capacity (see left part of Table~\ref{table_dataset}).
% After converting the RGB domain to YCbCr domain, watermarks are inserted into the Y-channel of given image for each MW method (\textbf{M1} $-$ \textbf{M7}).
% Likewise, watermark extraction is performed on the Y-channel of given image.

Discussing watermark embedding and extraction approach for color images, the process entails first converting images from the RGB to the YCbCr domain.
Watermarks are then inserted into and extracted from the Y-channel for each MW method (\textbf{M1} $-$ \textbf{M7}).
For each MW method, our WAN is trained with the hyperparameters $\lambda_{c}=0.4$ and $\lambda_{wa}=0.3$ during 50 epochs.
Except for the channel of the input image (i.e., $C=3$), the training settings are equal to the training methodology described in Section~\ref{sec_training_settings}.
In this experiment, the WAN is provided with an RGB image as input and is trained to invert a watermark bit that is inserted into a specific channel (i.e., the Y-channel) of a given image for each MW method.

\begin{figure*}[t]
\centering{\includegraphics[width=1.0\linewidth]{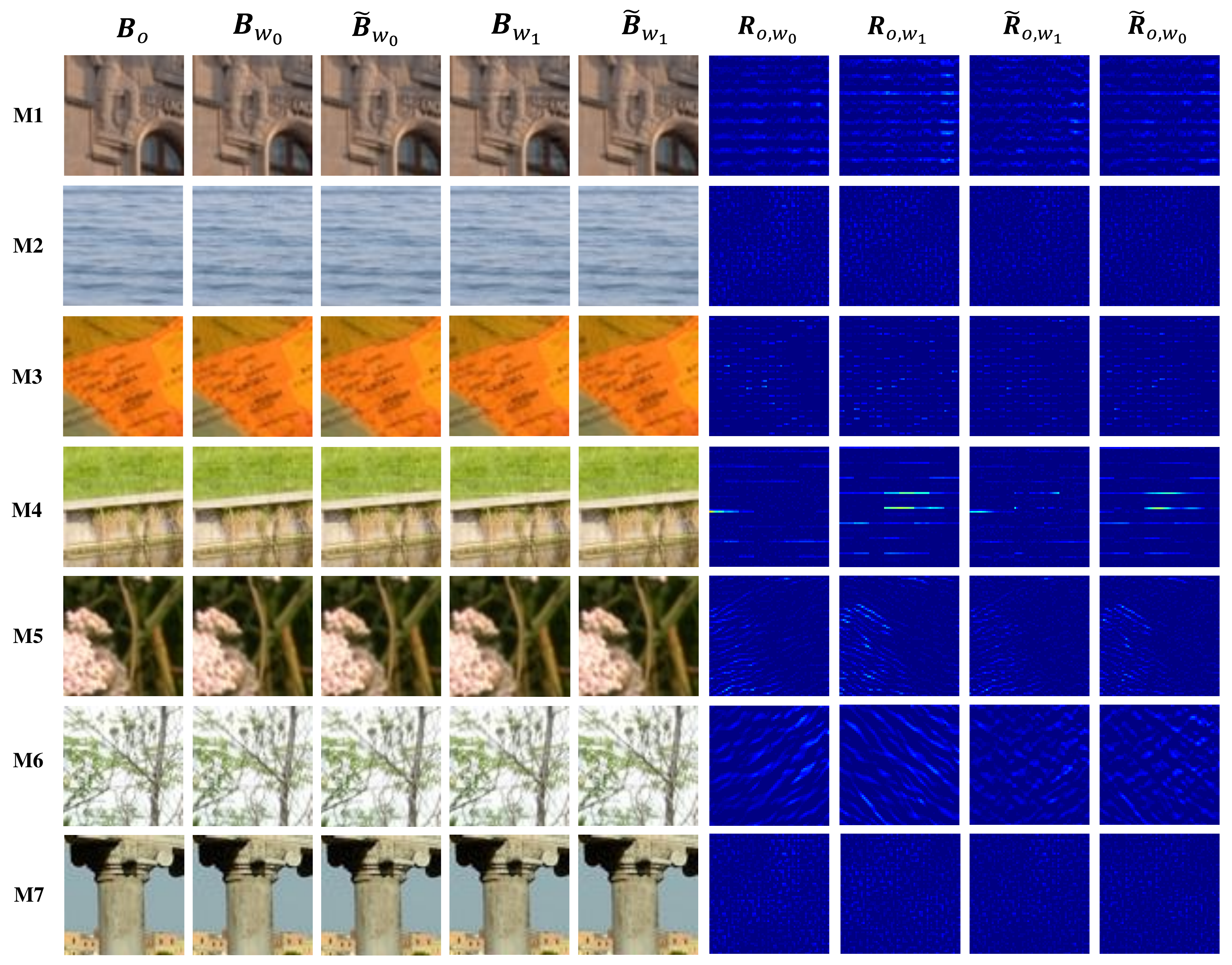}}
\caption{Qualitative evaluation results of the proposed WAN on color images with 1 bit of capacity}
\label{figure_results_color_image}
\end{figure*}

Table~\ref{table_results_color_image} presents the quantitative results of the WAN on a color image dataset.
In non-attack situation, each MW method has a low BER value (of 0.048 or less), while the average BER value increases dramatically to 0.916 after WAN is applied.
In case of color image, the average PSNR and SSIM values in non-attack situation are 41.51 dB and 0.973, respectively.
For average PSNR and SSIM, only minor degradations of 2.72 dB and 0.007, respectively, are shown after WAN-based attack.
In experiments with the color dataset, we confirm that effective bit inversion is achievable while reducing visual degradation, similar to observations with the base dataset (i.e., the gray-scale dataset).
It is expected that the performance difference from the results on grayscale images stems from the additional challenge of identifying the specific channel into which the watermark is inserted, without explicit guidance, in RGB format inputs.
Subsequently, the qualitative evaluation results of the proposed WAN on color images with a capacity of 1 bit are presented in Figure~\ref{figure_results_color_image}.
In the figure, upon close examination of the watermarked blocks and the blocks attacked as a result of the implementation of WAN, it can be observed that WAN hardly causes visual degradation.
The trends observed in the visualized residual data exhibit similarities to those found in the gray-scale dataset, as explored in Section~\ref{sec_qualitative_evaluation}.
% \color{red}
% Furthermore, as shown in attacked and residual images in Figure~\ref{figure_results_color_image}, it can be observed that WAN hardly causes visual degradation.
% It is expected that the performance difference from the results on gray-scale image is caused because the WAN should target the watermark signals inserted in a specific channel from the given RGB image.

\begin{table*}[t]
\caption{Quantitative evaluation results of the proposed AoW on fidelity and robustness}
% \scriptsize
\centering
\resizebox{\linewidth}{!}{
\begin{tabular}{lccccccccc}
\Xhline{2\arrayrulewidth}
\multirow{3}{*}{\shortstack{\textbf{MW method}}} & \multicolumn{3}{c}{\textbf{Watermarked}} & \multicolumn{3}{c}{\textbf{AoW-max}} & \multicolumn{3}{c}{\textbf{AoW-min}}\\
\cmidrule(r){2-4}\cmidrule(r){5-7}\cmidrule(r){8-10}
& PSNR & SSIM & BER & PSNR & SSIM & BER & PSNR & SSIM & BER\\
\hline
\textbf{M1} \cite{kim2018robust} & 35.55 & 0.938 & 0.026 & 31.67 & 0.914 & 0.020 & 40.01 & 0.979 & 0.085 \\
\textbf{M2} \cite{lin2011digital} & 41.86 & 0.988 & 0 & 36.70 & 0.973 & 0 & 43.49 & 0.992 & 0 \\
\textbf{M3} \cite{su2017improved} & 36.59 & 0.974 & 0 & 32.35 & 0.954 & 0 & 38.26 & 0.980 & 0\\
\textbf{M4} \cite{makbol2016block} & 38.98 & 0.986 & 0.002 & 36.01 & 0.978 & 0.002 & 41.87 & 0.991 & 0.009 \\
\textbf{M5} \cite{nam2020nsct} & 39.21 & 0.987 & 0.013 & 35.18 & 0.973 & 0.008 & 41.97 & 0.993 & 0.057 \\
\textbf{M6} \cite{kim2012dtcwt} & 36.61 & 0.972 & 0.021 & 33.52 & 0.957 & 0.016& 39.63 & 0.987 & 0.068\\
\textbf{M7} \cite{parah2016dct} & 36.43 & 0.945 & 0 & 34.09 & 0.936 & 0 & 43.54 & 0.986 & 0.082\\
\hline
Average & 37.89 & 0.970 & 0.009 & 34.21 & 0.955 & 0.007 & 41.25 & 0.987 & 0.043 \\
\Xhline{2\arrayrulewidth}
\end{tabular}
}
\label{table3}
\end{table*}

\subsection{Performance Evaluation of AoW}

\subsubsection{Quantitative Evaluation of AoW}

As introduced in the Section~\ref{sec_aow}, AoW is proposed to improve the performance in terms of imperceptibility or robustness of the target MW system.
The proposed AoW performs the role of an add-on module based on the resultant image from the inference of a pre-trained WAN for the target MW, without requiring separate training for AoW, as specified in Algorithm~\ref{algorithm1} and Equations (\ref{eq_aow_min}) and (\ref{eq_aow_max}).
For the quantitative evaluation of AoW, the original and watermarked images corresponding to the $64\times64$ resolution as specified on the left side of Table~\ref{table_dataset}, were utilized.
Residual data were calculated using the designated original and watermarked images to implement AoW-min or AoW-max for the target MW.
The performance of AoW is evaluated based on imperceptibility, assessed using PSNR and SSIM, and robustness, evaluated through BER.
In contrast to the WAN evaluation, which concentrated on factors leading to abnormal watermark extraction, AoW is assessed with a focus on enhancements in normal watermark extraction.

\begin{table*}[h]
\caption{Comparison of robustness of the AoW against signal processing attacks of each parameter}
% \scriptsize
\centering
\resizebox{\linewidth}{!}{
\begin{tabular}{cccccccccccccc}
\Xhline{2\arrayrulewidth}
\multirow{2}{*}{\textbf{Type}}& \multirow{3}{*}{\shortstack{\textbf{MW} \\\textbf{method}}} & \multicolumn{4}{c}{\textbf{JPEG}} & \multicolumn{4}{c}{\textbf{Median blur}} & \multicolumn{4}{c}{\textbf{Noise addition}}\\
\cmidrule(r){3-6}\cmidrule(r){7-10}\cmidrule(r){11-14}
& & 60 & 70 & 80 & Avg. & 2 & 3 & 4 & Avg. & 1 & 2 & 3 & Avg.\\
\hline
\multirow{7}{*}{WMed}& \textbf{M1} & 0.118 & 0.117 & 0.110 & 0.115 & 0.093 & 0.110 & 0.196 & 0.133 & 0.098 & 0.102 & 0.144 & 0.114 \\
& \textbf{M2} & 0 & 0 & 0 & 0 & 0.016 & 0.020 & 0.486 & 0.174 & 0 & 0 & 0.01 & 0.003 \\
& \textbf{M3} & 0.008 & 0.004 & 0 & 0.004 & 0.286 & 0.293 & 0.368 & 0.316 & 0 & 0.014 & 0.028 & 0.014 \\
& \textbf{M4} & 0.038 & 0.022 & 0.012 & 0.024 & 0.228 & 0.231 & 0.515 & 0.324 & 0.002 & 0.004 & 0.016 & 0.007 \\
& \textbf{M5} & 0.092 & 0.067 & 0.049 & 0.069 & 0.166 & 0.175 & 0.320 & 0.220 & 0.013 & 0.084 & 0.174 & 0.090 \\
& \textbf{M6} & 0.086 & 0.058 & 0.043 & 0.062 & 0.078 & 0.094 & 0.168 & 0.113 & 0.092 & 0.168 & 0.206 & 0.155 \\
& \textbf{M7} & 0.011 & 0.004 & 0 & 0.005 & 0.450 & 0.468 & 0.475 & 0.464 & 0 & 0.002 & 0.007 & 0.003 \\
\hline
\multirow{7}{*}{AoW-max}& \textbf{M1} & 0.066 & 0.070 & 0.062 & 0.066 & 0.071 & 0.075 & 0.144 & 0.096 & 0.046 & 0.056 & 0.128 & 0.76 \\
& \textbf{M2} & 0 & 0 & 0 & 0 & 0.010 & 0.013 & 0.423 & 0.148 & 0 & 0 & 0 & 0 \\
& \textbf{M3} & 0.007 & 0.003 & 0 & 0.003 & 0.285 & 0.288 & 0.364 & 0.312 & 0 & 0.002 & 0.010 & 0.004 \\
& \textbf{M4} & 0.034 & 0.018 & 0.012 & 0.021 & 0.192 & 0.218 & 0.496 & 0.302 & 0 & 0 & 0.014 & 0.004 \\
& \textbf{M5} & 0.086 & 0.065 & 0.044 & 0.065 & 0.152 & 0.173 & 0.306 & 0.210 & 0.013 & 0.079 & 0.170 & 0.087 \\
& \textbf{M6} & 0.044 & 0.029 & 0.024 & 0.032 & 0.064 & 0.60 & 0.144 & 0.089 & 0.042 & 0.110 & 0.182 & 0.111 \\
& \textbf{M7} & 0.010 & 0.001 & 0 & 0.004 & 0.412 & 0.424 & 0.443 & 0.426 & 0 & 0 & 0.002 & 0.001\\
\hline
\multirow{7}{*}{AoW-min}& \textbf{M1}  & 0.197 & 0.206 & 0.190 & 0.197 & 0.198 & 0.230 & 0.337 & 0.255 & 0.193 & 0.236 & 0.265 & 0.231\\
& \textbf{M2} & 0.006 & 0 & 0 & 0.002 & 0.036 & 0.038 & 0.475 & 0.183 & 0 & 0 & 0.040 & 0.013 \\
& \textbf{M3} & 0.018 & 0.006 & 0 & 0.008 & 0.288 & 0.297 & 0.373 & 0.319 & 0 & 0.012 & 0.018 & 0.010 \\
& \textbf{M4} & 0.092 & 0.056 & 0.035 & 0.061 & 0.263 & 0.286 & 0.517 & 0.355 & 0.010 & 0.034 & 0.055 & 0.033 \\
& \textbf{M5} & 0.170 & 0.135 & 0.122 & 0.142 & 0.246 & 0.280 & 0.356 & 0.294 & 0.066 & 0.168 & 0.242 & 0.158 \\
& \textbf{M6} & 0.271 & 0.260 & 0.252 & 0.261 & 0.208 & 0.228 & 0.282 & 0.239 & 0.251 & 0.272 & 0.378 & 0.300\\
& \textbf{M7} & 0.278 & 0.240 & 0.198 & 0.238 & 0.495 & 0.492 & 0.500 & 0.495 & 0.122 & 0.263 & 0.312 & 0.232\\
\Xhline{2\arrayrulewidth}
% \multicolumn{14}{l}{\emph{$\dagger$ JPEG, noise addition (NA), and median blur (MB) on the StirMark are used as watermarking attacks.}} \\ 
\multicolumn{14}{l}{\emph{$\dagger$ JPEG: [60, 70, 80], NA: [1, 2, 3], and MB: [2, 3, 4]}} \\ 
\end{tabular}
}
\label{table4}
\end{table*}

Table \ref{table3} reports the PSNR, SSIM, and BER of the original MW methods and their AoW-
max and AoW-min versions.
The AoW-max is introduced to improve robustness, and it achieves an average BER value of 0.007 across MW techniques \textbf{M1} to \textbf{M7}.
In addition, AoW-min achieves 41.25 dB in average PSNR, which is improved by 3.36 dB compared to the original MW methods, but the robustness of MW methods is slightly sacrificed.
Quantitative evaluation results specified in the table indicate that AoW-max and AoW-min each enable performance improvements in terms of robustness and imperceptibility, respectively. This trend is consistently observed across the diverse MW methods with various attributes.
Next, to evaluate the advantage of AoW-max, the robustness evaluation against signal processing attacks (i.e., JPEG, median blur, and noise addition) of StirMark is performed (see Table~\ref{table4}).
Here, parameters of JPEG, median blur, and noise addition are set to $[60,70,80]$, $[2,3,4]$, and $[1,2,3]$, respectively.
Although the average PSNR of AoW-max is 34.21 dB, which is degraded compared to original MW methods, the robustness against signal processing attacks is improved in terms of BER.
AoW-max shows lower BER than original methods as expected, and the improvements are more acquired in strong attacks than week attacks.
Specifically, notable improvements include a JPEG quality factor of 60, a median blur parameter of 4, and a noise addition parameter of 3.

\begin{table*}[t]
% \caption{Robustness evaluation of \textbf{M5} with AoW against geometric distortions}
\caption{Robustness Evaluation of MW Systems Designed for Resistance to Geometric Attacks with AoW}
\centering
\scriptsize
\begin{tabular}{clccc}
\Xhline{2\arrayrulewidth}
\multirow{2}{*}{\shortstack{\textbf{MW method}}} & \multirow{2}{*}{\shortstack{\textbf{Type}}} & \multicolumn{3}{c}{\textbf{Geometric distortion}}\\
\cmidrule(r){3-5}
& & Rotation & Center crop & Rescaling \\
\hline
& Watermarked & 0.106 & 0.055 & 0.047  \\
\textbf{M5} \cite{nam2020nsct} & AoW-max & 0.092 & 0.046 & 0.038  \\
& AoW-min & 0.185 & 0.086 & 0.093 \\
\hline
& Watermarked & 0.086 & 0.061 & 0.055  \\
\textbf{M6} \cite{kim2012dtcwt} & AoW-max & 0.079 & 0.044 &  0.047 \\
& AoW-min & 0.165 & 0.095 & 0.102 \\
\Xhline{2\arrayrulewidth}
\multicolumn{5}{l}{\emph{$\dagger$ Rotation: [3,4,5], Center crop: [85,90,95], Rescaling: [80,90,110,120]}} \\
% \multicolumn{4}{l}{\emph{$\dagger$ }} \\
\end{tabular}
\label{table_results_Aow_geo}
\end{table*}

Furthermore, we test the robustness against geometric distortions in Table~\ref{table_results_Aow_geo}.
As described in Equations~(\ref{eq_aow_min}) and (\ref{eq_aow_max}), since AoW operates based on residual data with original, watermarked, and attacked images, it is inherently influenced by the attributes of the target MW (e.g., robustness to geometric distortions).
Thus, we performed a quantitative evaluation on \textbf{M5} and \textbf{M6}, which are basically resistant to geometric distortions, and we verified efficacy of AoW-max against three types of attacks.
To verify the improvement in robustness against geometric attacks by AoW-max, robustness evaluation are conducted based on rotation, center crop, and rescaling in StirMark.
The parameters for rotation, center crop, and rescaling are set to $[3,4,5]$, $[85,90,95]$, and $[80,90,110,120]$, respectively.
We can find that AoW-max enhances the capability of \textbf{M5} and \textbf{M6} to withstand three types of geometric attacks.
We would like to note that the proposed AoW is applicable to assist in improving imperceptibility or robustness of MW methods, so watermarking designers can choose between AoW-max and AoW-min before redesigning watermarking systems to meet new requirements.

\begin{figure*}[t!]
\centering{\includegraphics[width=0.98\linewidth]{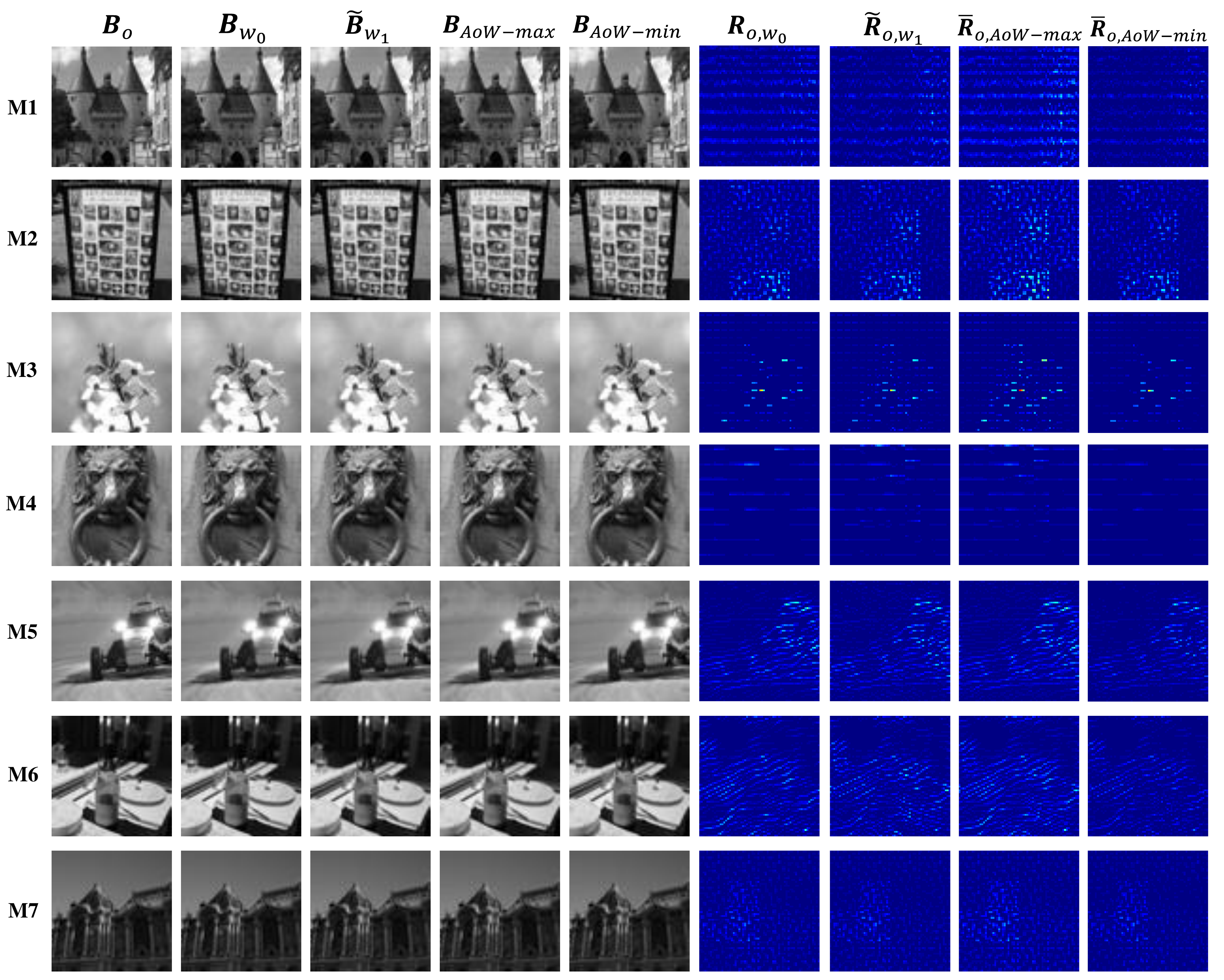}}
\caption{Qualitative evaluation results for AoW-min and AoW-max corresponding to case with watermark bit 0 embedded.}
\label{figure_qualitative_result_AoW_0}
\end{figure*}

\begin{figure*}[t!]
\centering{\includegraphics[width=0.98\linewidth]{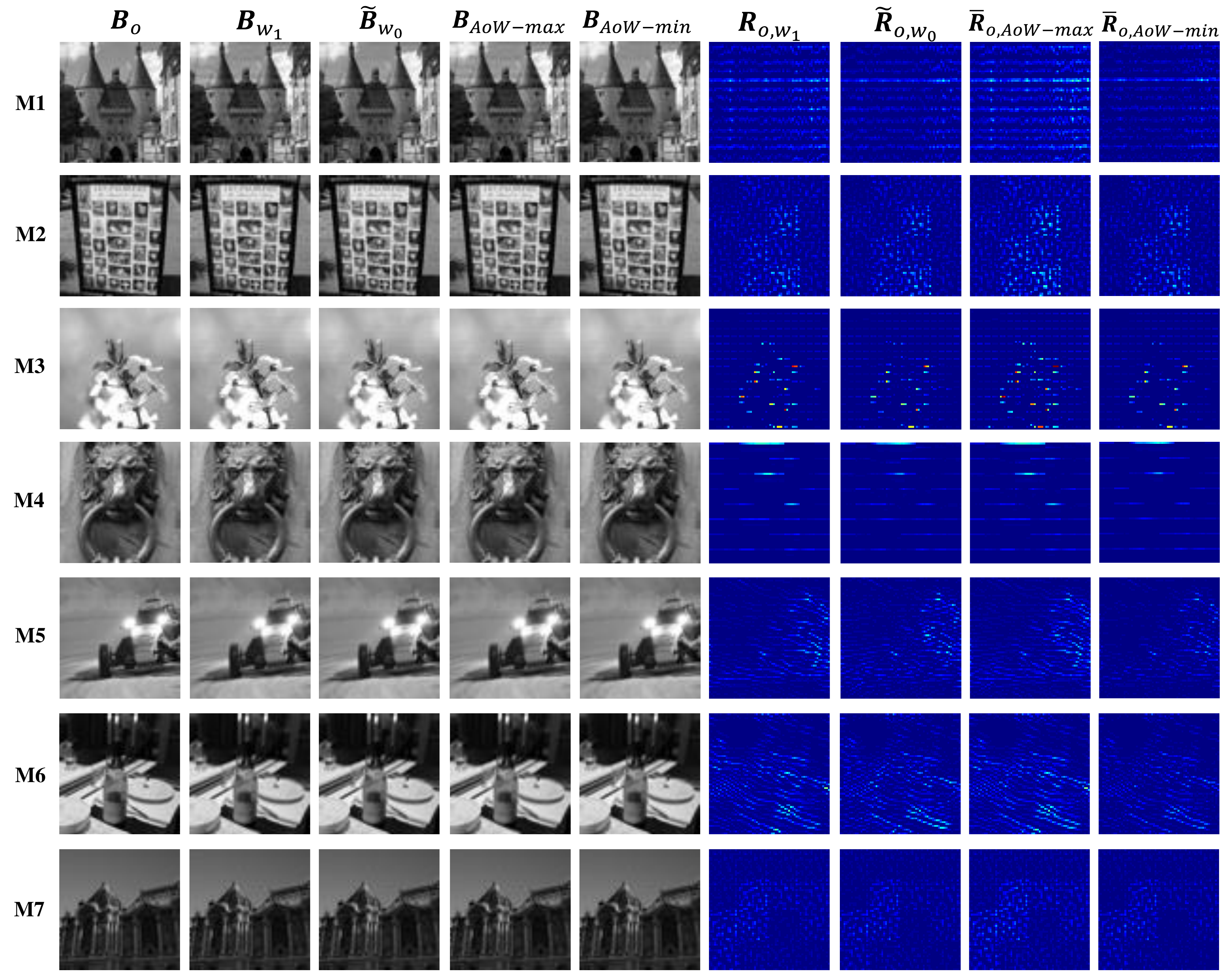}}
\caption{Qualitative evaluation results for AoW-min and AoW-max corresponding to case with watermark bit 1 embedded.}
\label{figure_qualitative_result_AoW_1}
\end{figure*}

\subsubsection{Qualitative Evaluation of AoW}

We also present the visualized results of residual images $(\bar{R}_{o,AoW-max}$ and $\bar{R}_{o,AoW-min}$) in Figure~\ref{figure_qualitative_result_AoW_0} and Figure~\ref{figure_qualitative_result_AoW_1}, where $\bar{R}_{o,AoW-max}$ and $\bar{R}_{o,AoW-min}$ are defined as ${B}_{o}-{B}_{AoW-max}$ and ${B}_{o}-{B}_{AoW-min}$, respectively.
As WAN inverts embedded bit successfully, we can find the $R_{o,w_{k}}$ and $\tilde{R}_{o,w_{k-1}}$ are similar for both bit ($k \in {0,1})$.
As shown in Figure~\ref{figure_qualitative_result_AoW_0} and Figure~\ref{figure_qualitative_result_AoW_1}, the signal intensity of residual images caused by AoW-min is relatively small compared to the others.
Although the patterns of residual are different according to watermarking methods, AoW-max shows more distinct patterns than the $R_{o,w_{k}}$, and $\overline{R}_{o,AoW-max}$ leaves less footprints in images than $R_{o,w_{k}}$.
Based on the results of quantitative and qualitative evaluation, we can confirm that AoW-min and AoW-max can be alternative to original watermarking methods for adjusting their imperceptibility and robustness.

\color{black}

\section{Conclusion}
\label{sec_Conclusion}

In this paper, we propose a novel neural network-based benchmark tool for block-based MW methods that exploits vulnerability of the targeted watermarking method and attacks watermarked images to mislead the watermarking extractor with minimal visual degradation.
To achieve this goal, we design customized losses of a watermarking attack loss for abnormal bit extraction and a content loss to maintain visual quality. 
Through quantitative and qualitative experiments with a variety of MW methods, we demonstrate that the WAN performs more effective attacks than existing benchmark tools in terms of maintaining visual quality and interfering with watermark extraction.
Furthermore, we show our WAN can be an add-on module, namely AoW, for existing MW methods to get additional performance gains.
Extensive experimental results show that AoW complements the performance of the targeted MW system by independently enhancing both imperceptibility and robustness.

\bibliographystyle{elsarticle-num} 
% \bibliography{reference}

\end{document}